\documentclass[aps,prd,reprint,amsmath,amssymb,superscriptaddress,floatfix]{revtex4}
\usepackage{amsmath}
\usepackage{graphicx}
\usepackage{subfigure}
\usepackage{amssymb}
\usepackage{xcolor}
\usepackage{multirow}
\usepackage{cancel}
\usepackage{color}
\usepackage{mathtools}
\usepackage{listings}
\usepackage{xcolor}
\usepackage{float}
\usepackage{slashed}
\usepackage{bm}
\usepackage[colorinlistoftodos]{todonotes}
\usepackage{diagbox}
\usepackage[colorlinks=true,citecolor=cyan,urlcolor=blue,bookmarks=true,bookmarks=true,bookmarksopen=true,bookmarksnumbered=true,bookmarksopenlevel=3]{hyperref}
\usepackage{subfigure}
\definecolor{airforceblue}{rgb}{0.36, 0.54, 0.66}
\definecolor{steelblue}{rgb}{0.27, 0.51, 0.71}
\definecolor{amber}{rgb}{1.0, 0.49, 0.0}

\newcommand{\be}{\begin{equation}}
\newcommand{\ee}{\end{equation}}
\newcommand{\beq}{\begin{equation}}
\newcommand{\eeq}{\end{equation}}
\newcommand{\bea}{\begin{eqnarray}}
\newcommand{\eea}{\end{eqnarray}}
\begin{document}

\title{Copositive criteria for a two-component dark matter model}
\author{\textsc{XinXin Qi}}
\author{\textsc{Hao Sun}\footnote{Corresponding author: haosun@dlut.edu.cn}}
\affiliation{ Institute of Theoretical Physics, School of Physics, Dalian University of Technology, \\ No.2 Linggong Road, Dalian, Liaoning, 116024, People’s Republic of China }
\date{\today}

\begin{abstract}
In this work, we investigate a two-component scalar dark matter framework featuring two singlet scalar fields as dark matter candidates. To ensure vacuum stability, we employ copositive criteria for the scalar potential. Guided by direct detection constraints, we analyze four distinct copositive scenarios characterized by specific negative parameter configurations. A comprehensive parameter space scan is performed under joint constraints from observed dark matter relic density and direct detection experiments.  Different signs of the couplings not only correspond to different copositive criteria but can also contribute to different parameter spaces due to the interference effect.  For the 4 different cases, the allowed values of quartic couplings are different but all demand that new Higgs plays a dominant role in determining dark matter relic density within the viable parameter space.
\vspace{0.5cm}
\end{abstract}
\maketitle
\setcounter{footnote}{0}

\section{Introduction}
\label{sec:intro}
  The Standard Model (SM) of particle physics has been remarkably successful in describing the fundamental particles and their interactions. However, the astrophysical observations indicate the existence of dark matter,
 which constitutes approximately $26.5\%$ of the universe's mass-energy content \cite{Planck:2015fie}. The existence of dark matter can be inferred from its gravitational effects on the universe's large-scale structure, yet its particle nature remains elusive. From the view of particle physics, dark matter can be assumed to be some stable particle carrying no SM gauge group quantum number.
 
One of the most popular and attractive dark matter candidates is weakly interacting massive particles(WIMPs)\cite{Qi:2021rhh,Qi:2022fzs,Qi:2021rpa}, where the observed dark matter relic density is generated via the so-called ``Freeze-out" mechanism\cite{Roszkowski:2017nbc}. Among the different WIMP scenarios, the singlet scalar dark matter model is the simplest where a new singlet scalar as dark matter is introduced to the SM. The singlet scalar is stabilized by extra $Z_2$ symmetry, and there are only two free model parameters in the model with dark matter mass and the quartic coupling of dark matter with SM Higgs. According to Higgs invisible decay search at LHC\cite{ATLAS:2016neq}, dark matter relic density, direct detection constraint \cite{He:2016mls,Escudero:2016gzx} and indirect detection constraint arising from gamma-ray spectrum induced by DM annihilation at Fermi-LAT \cite{Fermi-LAT:2015kyq,Fermi-LAT:2015att} and HESS\cite{HESS:2013rld}, the viable dark matter mass region is limited stringently with resonant mass being about half of the SM Higgs mass and high mass region larger than about 3 TeV \cite{Wu:2016mbe}. The current direct detection experiments have not yielded a clear signal, imposing stringent constraints on DM-nucleon interactions, so the WIMP scenarios are facing a serious crisis.

One possible solution to the problem is the multi-component dark matter models, which include two or more dark matter species. In the multi-component dark matter models, it is possible that one of the dark matter candidates has a large elastic cross-section on nuclei but forms only a subdominant DM component leading to a suppressed signal \cite{Belanger:2022qxt} therefore escapes from the direct detection constraint. There is no preference to assume that there is only one dark matter candidate in the universe, and discussion about multi-component dark matter can be found in \cite{Coleppa:2023vfh, Zurek:2008qg,Costa:2022lpy} and so on. Particularly, for the multi-component scalar dark matter model, one can introduce extra symmetry such as $Z_2 \times Z_2'$ \cite{Bhattacharya:2019fgs} , $Z_5$ symmetry \cite{Belanger:2020hyh,Qi:2024uiz,Qi:2023egb} ,$Z_7$ symmetry \cite{Belanger:2022esk} or other $Z_N$ symmetry \cite{Bhattacharya:2024nla,Yaguna:2019cvp,Yaguna:2021vhb} to ensure the stability of dark matter particles.

When constructing a viable multi-component scalar DM model, the stability of the electroweak (EW) vacuum should be guaranteed. For the scalar field theories, the concept of vacuum stability is intricately linked to the properties of the scalar potential, particularly its quartic couplings. A scalar potential that is bounded from below is a prerequisite for a stable vacuum, and this is where the copositive criteria come into play. Copositive matrices, a class of matrices that are positive on non-negative vectors, provide a mathematical framework to assess the stability conditions of scalar potentials. This approach allows for the derivation of analytic necessary and sufficient conditions for vacuum stability, which is essential for ensuring that the scalar potential does not lead to unphysical situations such as a vacuum that is not the global minimum of the potential. 

Discussion about copositive criteria and dark matter can be found in \cite{DuttaBanik:2020jrj,Kannike:2012pe,Chakrabortty:2013mha,Arhrib:2011uy,Ghorbani:2021rgs,Habibolahi:2022rcd}, and in this work, we consider a simple two-component scalar dark matter model and discuss the copositive criteria on the model. We introduce three singlet scalars $S_1$, $S_2$, and $S_3$ to the SM, where $S_1$ as well as $S_2$ as dark matter candidates and the singlet scalar $S_3$ has non-zero vacuum expectation value.  Note  that we come to the two component scalar dark matter model \cite{DiazSaez:2021pfw} when $S_3$ is absent, which demands a large $\lambda_{12}$ to evade direct detection and relic abundance constraints outside the Higgs resonance region. With the introduction of $S_3$, we have new annihilation channels and the couplings of DM with SM Higgs  as well as $\lambda_{12}$ can be smaller  to escape from the direct detection constraints with the observed DM relic density constraint. In other word, the existence of $S_3$ will contribute to the new dark matter annihilation processes so that we can have a wider parameter space compared with the dark matter models that only involve two singlet scalars. On the other hand, a $4\times 4$ copositive matrix of the quartic couplings can be constructed in the model. Note that the copositive criteria of the rank $\leqslant 3$  matrix can be given concretely, while for the higher rank, related discussions about the models are always ignored for their complexity. In this paper, we analyze the copositive criteria of the model systematically. Generally speaking, there are eight cases of the copositive criteria for the rank 4 matrix, and one can have more complex cases when considering the signs of the couplings of the model. Based on the direct detection constraint, we pick up 4 different cases from the point of exchange symmetry for simplicity and estimate the viable parameter space within the dark matter relic density constraint and direct detection constraint. The signs of the couplings will not only determine the copositive criteria but also make a difference in the cross-section of  $2 \to 2 $ processes related to the dark matter annihilation into other scalars due to the interference effect. We randomly scan the chosen parameter space and focus on the quartic couplings with $\lambda_{13}$,$\lambda_{23}$,$\lambda_{14}$ and $\lambda_{24}$. For the 4 different cases, the allowed values of $(|\lambda_{13}|,|\lambda_{14}|)$ are different but all demand that new Higgs plays a dominant role in determining dark matter relic density. Within the viable parameter space, copositive criteria can put stringent constraints as long as some coupling is negative. 

The paper is organized as follows. 
 We give the two-component scalar dark matter model in Sec.\ref{sec:2}.
In Sec.\ref{sec:3}, we discuss the copositivity of the symmetric matrix with eight different cases.
In Sec.\ref{sec:dm}, we consider the dark matter phenomenology of the model including relic density and direct detection.
In Sec.\ref{sec:so}, we  pick up 4 different cases by fixing some of the quartic parameters to be negative and discuss the effect of copositive constraint on the viable parameter space satisfying relic density and direct detection constraints.
Finally, we give a summary in Sec.\ref{sec:sum}.
\section{Model description}\label{sec:2}
In this section, we consider a simple two-component scalar dark matter model with $Z_2 \times Z_2' \times Z_2''$ symmetry by introducing three singlet scalars $S_{1,2,3}$, where the corresponding charges for dark matter particles $S_1$ and $S_2$ are $(-1,1,1)$ and $(1,-1,1)$ and $S_3$ has non-zero vacuum expectation value (vev) carrying $(1,1,-1)$ charge, while the SM particles carry $(1,1,1)$ charge. Therefore, the potential part of the scalars can be given by:
 \begin{align}
 \mathcal{V} &= \frac{1}{2}M_1^2 S_1^2 +  \frac{1}{2}M_2^2 S_2^2 - \frac{1}{2}\mu_3^2S_3^2 - \mu_H^2|H|^2 +\lambda_{H}|H|^4 + \frac{1}{4}\lambda_{11}S_1^4 +\frac{1}{4}\lambda_{22}S_2^4+  
 \frac{1}{4}\lambda_{33} S_3^4 \notag \\
 &+ \lambda_{12} S_1^2S_2^2 + \lambda_{13}S_1^2S_3^2 +\lambda_{14}S_1^2|H|^2 + \lambda_{23}S_2^2S_3^2 + \lambda_{24}S_2^2|H|^2  +  \lambda_{34}S_3^2|H|^2.
 \end{align}
  where $H$ is the SM Higgs doublet. Note that the $Z_2''$ symmetry is introduced to simplify the scalar potential in this work. Under unitarity gauge, $H$ and $S_3$ can be expressed with:
    \begin{equation}
H=\left(\begin{array}{c} 0 \\ \frac{v_0+h}{\sqrt{2}}\end{array} \right) \, , \quad
S_3=s_3+ v_1\, ,\quad
\end{equation}
 where $v_0 =246$ GeV corresponds to the electroweak symmetry breaking vev and $v_1$ is the vev of $S_3$. After spontaneous symmetry breaking (SSB), the masses of $S_1$ and $S_2$ can be given by:
 \begin{eqnarray}
 m_1^2= M_1^2 +2\lambda_{13}v_1^2 +\lambda_{14}v_0^2, m_2^2= M_2^2 +2                         \lambda_{23}v_1^2 +\lambda_{24}v_0^2.
 \end{eqnarray}
  where $m_1(m_2)$ represents the mass of $S_1(S_2)$. On the other hand, we have the squared mass matrix of $s_3$ and $h$ with:
  \begin{eqnarray}
    \mathcal{M}= \left(
    \begin{array}{cc}
     2\lambda_{33}v_1^2 & \lambda_{34}v_0v_1 \\
     \lambda_{34} v_0v_1 & 2\lambda_{H}v_0^2 \\
    \end{array}
    \right).
  \end{eqnarray}
The physical masses of the two Higgs states $h_1, h_2$ are then 
\begin{align}
\label{Higgsmass}
	m^2_{h_{1}} &= \lambda_H v_0^2 + \lambda_{33} v_1^2 
	- \sqrt{(\lambda_H v_0^2 - \lambda_{33} v_1^2)^2 + (\lambda_{34}v_0v_1)^2},\notag\\
  m^2_{h_{2}} &= \lambda_H v_0^2 + \lambda_{33} v_1^2 
	+ \sqrt{(\lambda_H v_0^2 - \lambda_{33} v_1^2)^2 + (\lambda_{34}v_0v_1)^2}.
\end{align} 
The mass eigenstate ($h_1,h_2)$ and the gauge eigenstate ($h, s_3$) can be related via
\begin{align}
\label{Higgs mixing}
	\begin{pmatrix}
		h_1 \\ h_2
	\end{pmatrix} = 
	\begin{pmatrix}
    	\cos\theta & -\sin\theta \\
		\sin\theta &  \cos\theta
	\end{pmatrix}
	\begin{pmatrix}
		h\\ s_3
	\end{pmatrix}.
\end{align} 
where
 \begin{eqnarray}
    \tan 2\theta= \frac{\lambda_{34}v_0v_1}{\lambda_{33}v_1^2 - \lambda_{H}v_0^2}.
\end{eqnarray}
Furthermore, we can assume $h_1$ is the observed SM Higgs and $h_2$ is the new Higgs in our model. One can choose the masses of the Higgs particles $m_{h_1}$ and $m_{h_2}$ as the inputs so that the couplings of $\lambda_H$, $\lambda_{33}$ and $\lambda_{34}$ can be given by:
\begin{align}
\label{para_quartic}
	\lambda_H &= 
 		\frac{(m_{h_1}^2 +m_{h_2}^2) - 
    	\cos 2 \theta (m_{h_2}^2 - m_{h_1}^2)}{4 v_0^2}, \nonumber\\
	\lambda_{33} &= 
 		\frac{(m_{h_1}^2 +m_{h_2}^2) + 
    	\cos 2 \theta (m_{h_2}^2 - m_{h_1}^2)}{4 v_1^2} , \\
	\lambda_{34} &= 
 		\frac{\sin 2 \theta (m_{h_2}^2 - m_{h_1}^2)}{2 v_0 v_1}.  \nonumber
\end{align} 
\section{ Copositivity of symmetric matrix}
\label{sec:3}
 The extended scalar field models beyond the SM can contain complicated quartic
couplings, thus it is important to formulate the criteria that
define the boundedness of the potential with the largest parameter space from the view of the vacuum stability constraint. One possible idea is to construct the quartic couplings
as a pure square of the combinations of bilinear scalar fields
and set their coefficients to be non-negative.  Here we have imported the idea of copositivity of symmetric matrices. In our model, the scalar potential quartic terms can be given with a symmetric matrix as follows:
\begin{eqnarray}\label{m1}
  \mathcal{S}=\left(
  \begin{array}{cccc}
    \lambda_{11}&\lambda_{12}&\lambda_{13}&\lambda_{14}\\
     &\lambda_{22}&\lambda_{23}&\lambda_{24}\\
     &&\lambda_{33}&\lambda_{34}\\
     &&&\lambda_{44}
  \end{array}
      \right).
\end{eqnarray}
In the following discussion, we use $\lambda_{44}$  to represent $\lambda_H$ to make our discussion more coherent.
For this symmetric matrix of order four, we should convict eight different cases depending on the sign distributions of the off-diagonal elements to determine whether this matrix is copositive. In these cases, all the diagonal elements should be positive as a generic condition, and the eight cases are given as follows \cite{Chakrabortty:2013mha}:
\begin{itemize}
\item  Case I:  The matrix $\mathcal{S}$ is copositive if and only if $\lambda_{ii} \geq 0$ when all the off-diagonal elements are positive.\\
\item Case II:  The matrix $\mathcal{S}$ is copositive if and
only if $(\lambda_{ii}\lambda_{jj}-\lambda_{ij}^2) \geq 0$ when  $\lambda_{ij} \leq 0$ and other off-diagonal elements are positive.\\
\item Case III:  The matrix $\mathcal{S}$ is copositive  if and only if $(\lambda_{ii}\lambda_{jj}-\lambda_{ij}^2)\geq 0, (\lambda_{ll}\lambda_{kk}-\lambda_{lk}^2) \geq 0$ when $\lambda_{ij},\lambda_{lk} \leq 0$  and other off-diagonal elements are positive.\\
\item Case IV:  When $\lambda_{ij},\lambda_{ik} \leq 0$,  we should have $(\lambda_{ii}\lambda_{jk}-\lambda_{ij}\lambda_{ik}+\sqrt{(\lambda_{ii}\lambda_{jj}-\lambda_{ij}^2)(\lambda_{ii}\lambda_{kk}-\lambda_{ik}^2)}) \\
\geq 0$ to make sure 
the matrix $\mathcal{S}$ is copositive.\\
\item Case V:   The matrix $\mathcal{S}$ is copositive  if and only if the following order three
matrix is copositive:
\be \nonumber \left( 
		\begin{array}{ccc}
                \lambda_{ii}  & \lambda_{ij} & \lambda_{ik}\\ 
                 & \lambda_{jj} & \lambda_{jk}\\ 
                 & & \lambda_{kk}\\ 
               \end{array}
               \right).
\ee     
 if $\lambda_{ij},\lambda_{jk},\lambda_{ik}\leq 0$ while the other off-diagonal elements are positive.\\
\item  Case VI:  The matrix $\mathcal{S}$ is copositive  if the following matrix is copositive:
\be  \nonumber \left( 
		\begin{array}{ccc}
                \lambda_{ii}\lambda_{jj}-\lambda_{ij}^2  & \lambda_{ii}\lambda_{jk}-\lambda_{ij}\lambda_{ik} & \lambda_{ii}\lambda_{jl}-\lambda_{ij}\lambda_{il}\\
                 & \lambda_{ii}\lambda_{kk}-\lambda_{ik}^2 & \lambda_{ii}\lambda_{kl}-\lambda_{ik}\lambda_{il}\\
                & & \lambda_{ii}\lambda_{ll}-\lambda_{il}^2\\
               \end{array}
               \right).
\ee 
when $\lambda_{ij},\lambda_{ik},\lambda_{il}\leq 0$ and other off-diagonal elements are positive.\\
\item Case VII: The following matrix of order three should be copositive to make sure $\mathcal{S}$ is copositive:
\be  \nonumber \left( 
\begin{array}{ccc}
\lambda_{kk} \big(\lambda_{jj}\lambda_{ik}^2-2\lambda_{ij}\lambda_{ik}\lambda_{jk}+ \lambda_{ii} \lambda_{jk}^2\big)  & \lambda_{kk} (\lambda_{jj}\lambda_{ik} - \lambda_{ij} \lambda_{jk}) & \lambda_{kk}(\lambda_{ik}\lambda_{jl} - \lambda_{jk} \lambda_{il})\\  
 & \lambda_{jj}\lambda_{kk} - \lambda_{jk}^2 & \lambda_{kk}\lambda_{jl} - \lambda_{jk} \lambda_{kl}\\
  &  & \lambda_{kk}\lambda_{ll} - \lambda_{kl}^2\\ 
\end{array}
\right).
\ee 
when $\lambda_{ij},\lambda_{jk},\lambda_{kl}\leq 0$ and other off-diagonal elements are positive.\\
\item Case VIII: The following matrix of order three should be copositive to make sure $\mathcal{S}$ is copositive:
\be  \nonumber \left( 
		\begin{array}{ccc}
                \lambda_{ll} (\lambda_{ii}\lambda_{jl}^2-2\lambda_{ij}\lambda_{il}\lambda_{jl}+ \lambda_{jj} \lambda_{il}^2)  &  \lambda_{ll} (\lambda_{ii}\lambda_{jl} - \lambda_{ij} \lambda_{il}) & \lambda_{ll}(\lambda_{ik}\lambda_{jl} - \lambda_{il} \lambda_{jk})\\ 
                 & \lambda_{ii}\lambda_{ll} - \lambda_{il}^2 & \lambda_{ll}\lambda_{ik} - \lambda_{il} \lambda_{kl}\\ 
                &  & \lambda_{kk}\lambda_{ll} - \lambda_{kl}^2\\ 
               \end{array} 
               \right).
\ee 
 when $\lambda_{ij},\lambda_{jk},\lambda_{kl}, \lambda_{il} \leq 0$ and other off-diagonal elements are positive.
\end{itemize}
We should discuss these eight cases separately and thus will have different allowed parameter space from the view of theoretical constraint.

\section{ Dark matter phenomenology}\label{sec:dm}
\begin{figure}[htbp]
\centering
 \includegraphics[height=12cm,width=12cm]{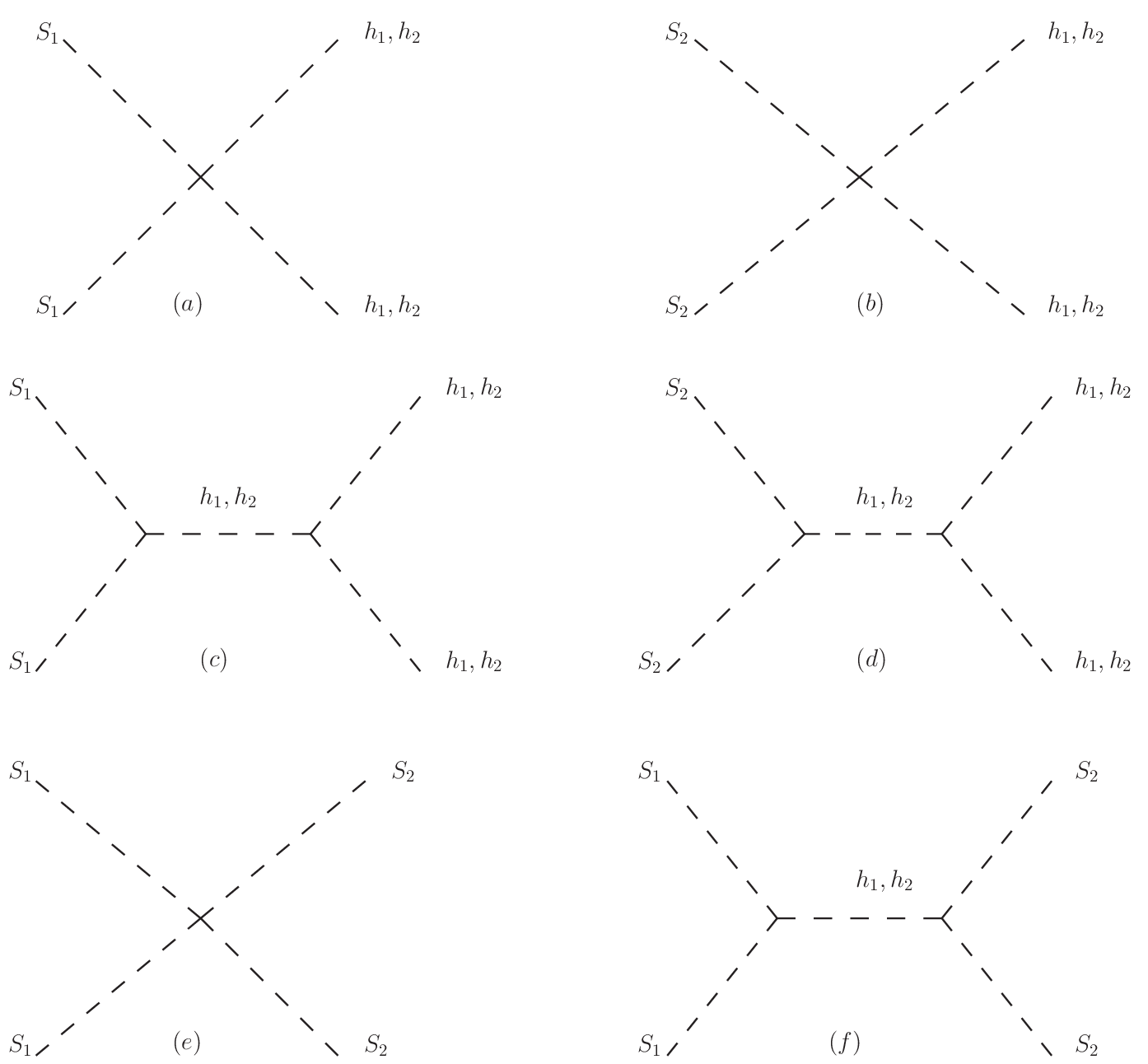}
 \caption{Feynman diagrams of dark matter annihilation processes related to new particles.}
 \label{feyn}
 \end{figure}

 In our model, there are two scalar dark matter particles stabilized by $Z_2 \times Z_2'\times Z_2''$ symmetry, which are related to SM particles with Higgs-portal interactions. The current observed dark matter relic density given by the Planck collaboration is $\Omega_{DM}h^2 = 0.1198 \pm 0.0012$ \cite{Belanger:2020gnr}, and we consider dark matter production in our model to be generated with the ``Freeze-out" mechanism. Both $S_1$ and $S_2$ will contribute to dark matter relic density and the Boltzmann equations for the number density of $S_1$
  and $S_2$ are given as follows:
  \begin{eqnarray} \nonumber
\frac{dn_1}{dt} + 3Hn_1 &=& - \langle \sigma v \rangle^{S_1S_1\to XX}(n_1^2-{\bar{n_1}}^2)
                        - \langle \sigma v \rangle^{S_1S_1\to h_{1,2}h_{1,2}}(n_1^2-{\bar{n_1}}^2) \\
		        &-&  \langle \sigma v \rangle^{ S_1S_1 \to S_2 S_2}(n_1^2-n_2^2 \frac{\bar{n_1}^2}{\bar{n_2}^2}), \notag
\end{eqnarray}
\begin{eqnarray}
 \frac{dn_2}{dt} + 3Hn_2 &=& - \langle \sigma v \rangle^{S_2S_2\to XX}(n_2^2-{\bar{n_2}}^2)
                        - \langle \sigma v \rangle^{S_2S_2\to h_{1,2}h_{1,2}}(n_2^2-{\bar{n_2}}^2) \notag \\
		        &-&  \langle \sigma v \rangle^{ S_2S_2 \to S_1 S_1}(n_2^2-n_1^2 \frac{\bar{n_2}^2}{\bar{n_1}^2}). \ \ \ \         
\end{eqnarray}
where $n_1(n_2)$ is the number density of $S_1(S_2)$, $\bar{n_1}(\bar{n_2})$ is the number density in thermal equilibrium, $H$ is the Hubble expansion rate of the Universe, $X$ denotes SM particles and $\langle \sigma v \rangle$ is the thermally averaged annihilation cross section. In Fig.~\ref{feyn}, we show the Feynman diagrams of dark matter annihilation processes related to new particles. Note that both $S_1$ and $S_2$ can be dominant in DM relic density due to the exchange symmetry, which depends on the mass heirarchy between $S_1$ and $S_2$, and the lighter one will be dominant since heavy component can annihilate into the lighter component.
To calculate the DM relic density numerically we use the micrOMGEAs 5.0.6 package \cite{Alguero:2022inz}, in which the model has been implemented through the FeynRules package \cite{Alloul:2013bka}.

On the other hand, dark matter can scatter off the nuclei with the Higgs-mediated t-channel processes, which should be constrained by direct detection results. The effective Lagrangian related to DM-quark elastic scattering can be given by \cite{Basak:2021tnj}:
\begin{eqnarray}
 \mathcal{L}_{q,eff}=\sum_{S=S_1,S_2} -\frac{m_q}{2v_0}(\frac{\mathcal{C}_{h_1SS}}{m_{h_1}^2}+\frac{\mathcal{C}_{h_2SS}}{m_{h_2}^2})SS\bar{q}q.
 \end{eqnarray}
  where $\mathcal{C}_{h_1SS} $ as well as $\mathcal{C}_{h_2SS}$ are the couplings of dark matter $S_{1,2}$ with the $h_{1,2}$ and
  \begin{eqnarray}
      \mathcal{C}_{h_1S_1S_1}=-2 \cos\theta \lambda_{14}v_0 +2\sin\theta \lambda_{13}v_1, \notag\\
     \mathcal{C}_{h_2S_1S_1}=-2 \sin\theta \lambda_{14}v_0 -2\cos\theta \lambda_{13}v_1, \notag\\  
      \mathcal{C}_{h_1S_2S_2}=-2 \cos\theta \lambda_{24}v_0 +2\sin\theta \lambda_{23}v_1, \notag\\ 
       \mathcal{C}_{h_2S_2S_2}=-2 \sin\theta \lambda_{24}v_0 -2\cos\theta \lambda_{23}v_1.
  \end{eqnarray}
The expression  of  the cross-section for the spin-independent DM- nucleon elastic scattering  can thus  be given as follows \cite{He:2008qm}:
\begin{eqnarray}
    \sigma_{S_iN } =\frac{m_N^4f_N^2}{4\pi(m_N+m_i)^2} \times (\frac{\mathcal{C}_{h_1S_iS_i}}{m_{h_1}^2}\cos\theta+\frac{\mathcal{C}_{h_2S_iS_i}}{m_{h_2}^2}\sin\theta)^2 ,(i=1,2)
\end{eqnarray}
 where $m_N$ represents the nucleon mass and  $f_N$ is the Higgs-nucleon Form factor with $f_N=0.308(18)$ according to the 
phenomenological and lattice-QCD calculations \cite{Bell:2013wua}.
Since we have two dark matter particles in the model, we should compare the value of $\xi_i \sigma_{S_i N }$ against the direct detection constraints provided by the experiment results instead of the cross-section, where $\xi_i$ is the fraction of $S_i$ in the total relic density defined by:
\begin{eqnarray}
    \xi_i =\frac{\Omega_i}{\Omega_{DM}}, (i=1,2).
\end{eqnarray}
and $\Omega_{DM}$ is the observed value of the dark matter density $\Omega_i$ is the density of $S_i(i=1,2)$.  In our analysis, we will consider the current direct detection limit set by the LZ result \cite{LZ:2024zvo}. 

\section{Discussion}\label{sec:so}
In this part, we estimate the parameter space from the view of copositive criteria and dark matter phenomenology related to relic density and direct detection constraints. Provided that direct detection constraint is more stringent on the low mass region, we focus on the TeV scale dark matter mass in this work, where indirect detection constraints make little difference on the parameter space.
 As we mentioned above, there are eight cases related to copositive criteria that depend on the sign of the parameters, and each case corresponds to a viable parameter space that should be limited by the relic density constraint and direct detection constraint. On the other hand, regardless of the symmetric matrix $\mathcal{S}$ with the permutation $\lambda_{ij}=\lambda_{ji}$,  we can still have different results when the indexes $i,j,k,l$ take different permutations even just for one case. Hence, discussion about the parameter space will be complex for different permutations of the indexes and one should consider all the possible cases to obtain the viable parameter space. 
\subsection{Estimate on the parameters}
 Before discussion, we can estimate the parameters to simplify the discussion.  In this work,  we  choose the following parameters as inputs:
\begin{eqnarray}
v_1, \sin\theta ,m_{h_2}, m_{1,2},                                                                                                                                                                                                                                                                                                                                                                                                                                                                                                                                                                                                                                                              
\lambda_{11},\lambda_{12},\lambda_{13},\lambda_{14},\lambda_{22},\lambda_{23},\lambda_{24}.
\end{eqnarray}
while $\lambda_{33}$,$\lambda_{34}$ and $\lambda_{44}(\lambda_H)$ can be given by Eq.~\ref{para_quartic}. Note that $\lambda_{34}$ is the coupling of the new Higgs particles, which makes little difference in dark matter results, and one can fix $v_1$, $\sin\theta$ as well as $m_{h_2}$ so that $\lambda_{33},\lambda_{34}$ and $\lambda_{44}(\lambda_H)$ can be given explicitly.  Note that the self-interaction couplings $\lambda_{11}$ and $\lambda_{22}$ make little difference in dark matter production, and we fix $\lambda_{11}=\lambda_{22}=\pi$ for simplicity. Therefore, we can just consider the signs of $\lambda_{12},\lambda_{13},\lambda_{23},\lambda_{14},\lambda_{24}$. Notice that we have an exchange symmetry of $\lambda_{13} \leftrightarrow \lambda_{23}$, $\lambda_{14} \leftrightarrow \lambda_{24}$ and $S_1 \leftrightarrow S_2$ due to the $Z_2 \times Z_2'\times Z_2''$ symmetry and there is no preference for the permutations of the indexes, one can group the five couplings into $\lambda_{12}$,$(\lambda_{13},\lambda_{14})$ and $(\lambda_{23},\lambda_{24})$. 

\begin{figure}[htbp]
\centering
 \includegraphics[height=6.5cm,width=7cm]{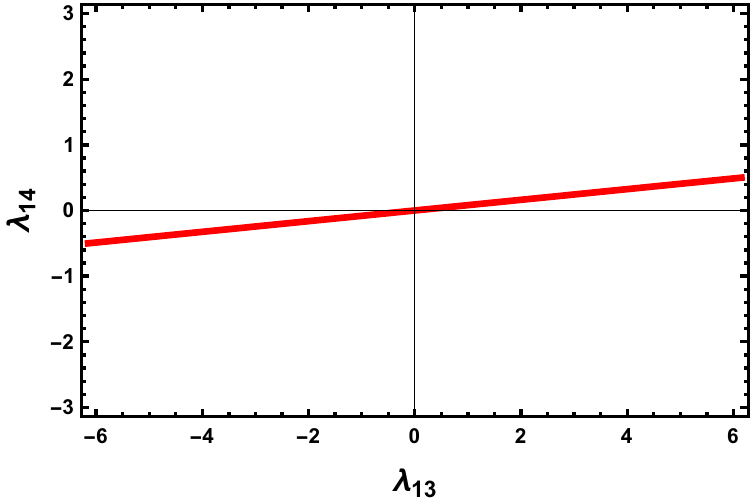}
 \caption{Viable parameter space of $(\lambda_{13},\lambda_{14})$ under direct detection constraint in the absence of $S_2$ with $m_1= 1$ TeV.}
 \label{dd}
 \end{figure}

The direct detection result gives the most stringent constraint on the parameter space. In Fig.~\ref{dd}, we give the viable parameter space of $(\lambda_{13},\lambda_{14})$ satisfying direct detection constraint with  $m_1= 1$ TeV in the absence of $S_2$.  $\lambda_{14}$, the coupling of SM Higgs with $S_1$ is limited within a narrow region under direct detection constraint while the value of $\lambda_{13}$ is more flexible.
 For the heavier $m_1$,  one can have similar allowed region for $(\lambda_{13},\lambda_{14})$ according to the expression of the DM- nucleon elastic scattering.
On the other hand,  a negative(positive) $\lambda_{13}$ always demands a negative(positive) $\lambda_{14}$ for large $|\lambda_{13}|$, while the small $|\lambda_{13}|$ and $|\lambda_{14}|$ will induce dark matter production over-abundant even though the parameter space is under direct detection constraint. Therefore, one can conclude that $\lambda_{13}$ and $\lambda_{14}$ taking the same signs under direct detection constraint, which is also true for $\lambda_{23}$ and $\lambda_{24}$ due to the exchange symmetry. Moreover,  we are left with $7(=2^3-1)$ cases to be discussed, where the case $\lambda_{12},\lambda_{13},\lambda_{14},\lambda_{23}$, $\lambda_{24}\leqslant 0$ is excluded according to the copositive criteria. Due to the exchange symmetry, the parameter space of $(\lambda_{13},\lambda_{14})$ and $(\lambda_{23},\lambda_{24})$  can be similar, we hence just consider one case to decrease the complexity and simplify the discussion. Concretely speaking,
 we will just focus on the viable parameter space of  $(\lambda_{13},\lambda_{14})$ but fix $\lambda_{23}$ and $\lambda_{24}$  to be constants in the following discussion. We have 5 cases left to discuss with (1)$\lambda_{12},\lambda_{13},\lambda_{14},\lambda_{23},\lambda_{24} \geqslant 0$. (2)$\lambda_{12},\lambda_{23},\lambda_{24}\geqslant 0, \lambda_{13},\lambda_{14}\leqslant 0$. (3)$\lambda_{12}\geqslant 0,\lambda_{13},\lambda_{14},\lambda_{23},\lambda_{24}\leqslant 0$. (4)$\lambda_{12},\lambda_{13},\lambda_{14}\leqslant 0, \lambda_{23},\lambda_{24} \geqslant 0$. (5)$\lambda_{12}\leqslant 0, \lambda_{13},\lambda_{14},\lambda_{23},\lambda_{24}\geqslant 0$.  Note that case (5) always satisfies copositive criteria since the inequality $\lambda_{11}\lambda_{22} -\lambda_{12}^2 \geqslant 0$ always holds, and there is no difference between case (1) and case (5) when considering copositive criteria. Therefore, we ignore case (5) and focus on the first four cases in the following discussion.

 \subsection{Theortical constraint of copositive criteria}
Following the above discussion, we have 4 different cases left with the theoretical constraints of copositive criteria to discuss. The case (1) corresponds to Case I  in Sec~\ref{sec:3} where all the couplings are larger than zero. The case (2) corresponds to Case IV with $\lambda_{13},\lambda_{14}\leqslant 0$, and the copositive criterial demands that:
\begin{eqnarray}
 \lambda_{11}\lambda_{34}-\lambda_{13}\lambda_{14}+\sqrt{(\lambda_{11}\lambda_{33}-\lambda_{13}^2)(\lambda_{11}\lambda_{44}-\lambda_{14}^2)} \geqslant 0,
\end{eqnarray}
The case (3) correspond to Case VI with $\lambda_{12},\lambda_{13},\lambda_{14}\leqslant 0$,  and the copositive criteria demands that:
\begin{eqnarray}
\lambda_{11}\lambda_{23}-\lambda_{12}\lambda_{13} + \sqrt{(\lambda_{11}\lambda_{22}-\lambda_{12}^2)(\lambda_{11}\lambda_{33}-\lambda_{13}^2)} \geqslant 0,
 \label{51}
\end{eqnarray}
\begin{eqnarray}
\lambda_{11}\lambda_{24}-\lambda_{12}\lambda_{14} + \sqrt{(\lambda_{11}\lambda_{22}-\lambda_{12}^2)(\lambda_{11}\lambda_{44}-\lambda_{14}^2)} \geqslant 0,
\label{52}
\end{eqnarray} 
\begin{eqnarray}
\lambda_{11}\lambda_{34}-\lambda_{13}\lambda_{14} + \sqrt{(\lambda_{11}\lambda_{33}-\lambda_{13}^2)(\lambda_{11}\lambda_{44}-\lambda_{14}^2)} \geqslant 0, 
\label{53}
\end{eqnarray}
\begin{eqnarray*}
\sqrt{2}\sqrt{(\lambda_{11}\lambda_{24}-\lambda_{12}\lambda_{14})+\sqrt{(\lambda_{11}\lambda_{22}-\lambda_{12}^2)
(\lambda_{11}\lambda_{44}-\lambda_{14}^2)}}\sqrt{(\lambda_{11}\lambda_{23}-\lambda_{12}\lambda_{13})+\sqrt{(\lambda_{11}\lambda_{22}-\lambda_{12}^2)(\lambda_{11}\lambda_{33}-\lambda_{13}^2)}}
\end{eqnarray*}
\begin{eqnarray*}
\sqrt{(\lambda_{11}\lambda_{34}-\lambda_{13}\lambda_{14})+\sqrt{(\lambda_{11}\lambda_{33}-\lambda_{13}^2)(\lambda_{11}\lambda_{44}-\lambda_{14}^2)}} + \sqrt{(\lambda_{11}\lambda_{22}-\lambda_{12}^2)(\lambda_{11}\lambda_{33}-\lambda_{13}^2)(\lambda_{11}\lambda_{44}-\lambda_{14}^2)}+(\lambda_{11}\lambda_{23}
\label{54}
\end{eqnarray*}
\begin{eqnarray}
 -\lambda_{12}\lambda_{13})\sqrt{\lambda_{11}\lambda_{44}-\lambda_{14}^2}
+(\lambda_{11}\lambda_{24}-\lambda_{12}\lambda_{14})\sqrt{\lambda_{11}\lambda_{33}-\lambda_{13}^2} + (\lambda_{11}\lambda_{34}-\lambda_{13}\lambda_{14})\sqrt{\lambda_{11}\lambda_{22}-\lambda_{12}^2}\geqslant 0.
\end{eqnarray}

The case (4) belongs to Case VIII with $\lambda_{12}\geqslant 0, \lambda_{13},\lambda_{14},\lambda_{23},\lambda_{24}\leqslant 0$, one can define:
\begin{eqnarray*}
A_{11}= \lambda_{22}(\lambda_{33}\lambda_{12}^2-2\lambda_{13}\lambda_{23}\lambda_{12}+\lambda_{11}\lambda_{23}^2),A_{12}=\lambda_{22}(\lambda_{33}\lambda_{12}-\lambda_{13}\lambda_{23}),A_{13}=\lambda_{22}(\lambda_{34}\lambda_{12}-\lambda_{23}\lambda_{14}),
\end{eqnarray*}
\begin{eqnarray*}
A_{22}=\lambda_{33}\lambda_{22}-\lambda_{23}^2, A_{23}=\lambda_{22}\lambda_{34}-\lambda_{23}\lambda_{24}, A_{33}=\lambda_{44}\lambda_{22}-\lambda_{24}^2.
\end{eqnarray*}
Therefore, the copositive criteria demands that:
\begin{eqnarray}\label{eq21}
A_{11} \geqslant 0,A_{22}\geqslant 0, A_{33}\geqslant 0,
A_{12}+ \sqrt{A_{11}A_{22}} \geqslant 0,
A_{13} +\sqrt{A_{11}A_{33}} \geqslant 0,
 A_{23} +\sqrt{A_{22}A_{33}} \geqslant 0,
\label{515}
\end{eqnarray}
\begin{align}
&A_{13}\sqrt{A_{22}}+A_{23}\sqrt{A_{11}}+\sqrt{2(A_{12}+\sqrt{A_{11}A_{22}})(A_{13}+\sqrt{A_{11}A_{33}})
(A_{23}+\sqrt{A_{22}A_{33}})} +\sqrt{A_{11}A_{22}A_{33}} +A_{12}\sqrt{A_{33}} \geqslant 0.
\label{516}
\end{align}
As we mentioned above, to obtain a stable vacuum, copositive criteria can put stringent constraints on the parameter space, and different copositive criteria will contribute to different viable parameter space which depends on the signs of the quartic couplings.

\subsection{Results}
In this part, we discuss the copositive criteria, dark matter relic density and direct detection constraint on the model. We consider the four different cases of the copositive criteria and focus on the viable parameter space of $(\lambda_{13},\lambda_{14})$. We fix $v_1=2$ TeV, $m_{h_2}=1.5$ TeV and $\sin\theta=0.01$ so that we can obtain $\lambda_{33}=0.2812$, $\lambda_{34}=0.0454$ and $\lambda_{44}=0.1309$. In addition, we set $|\lambda_{12}|=1$, $m_2=3.5$ TeV and fix $\lambda_{23}$ as well as $\lambda_{24}$ constants. Note that we have chosen the small $\sin\theta$ so that the contribution of $S_iS_i \to h_1h_2 (i=1,2)$ is suppressed. The expression of the processes related to dark matter production in the limit of $\sin\theta \to 0$ can be found in App.~\ref{csa}. We randomly scan the parameter space  with
 \begin{eqnarray*}
  m_1 \subset \{1~\mathrm{TeV}, 2~\mathrm{TeV}, 3~\mathrm{TeV}\},
  |\lambda_{13}| \subset [0.0001,3.14],
   |\lambda_{14}| \subset [0.0001,3.14].
 \end{eqnarray*}
 In our scans, we consider the model compatible with the observed dark matter relic density if the value, as given by micrOMEGAs, lies between 0.119 and 0.121.
 
\begin{figure}[htbp]
\centering
 \subfigure[]{\includegraphics[height=6cm,width=6cm]{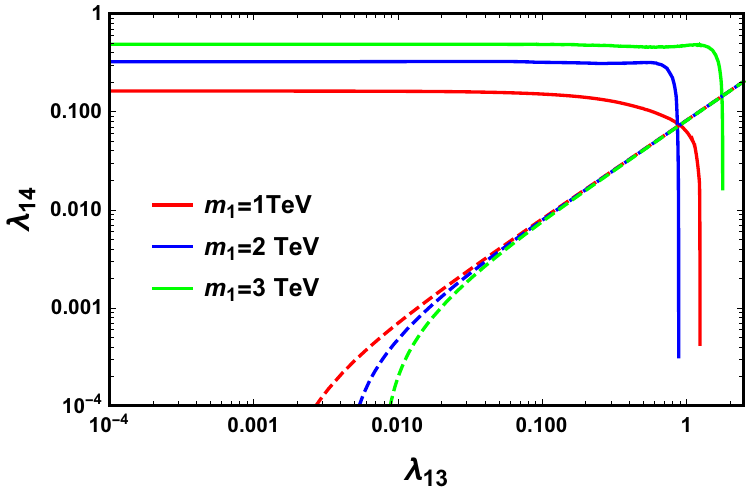}}
\subfigure[]{\includegraphics[height=6cm,width=6cm]{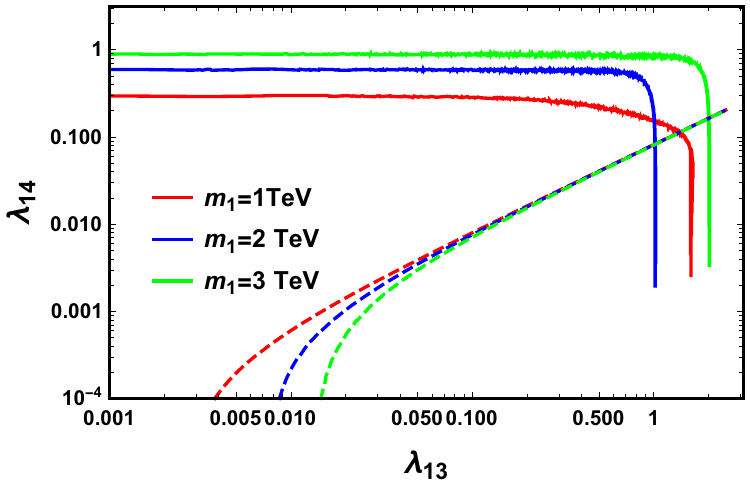}}
\caption{Results of case (1) with $\lambda_{12}=1,\lambda_{13},\lambda_{14} \geqslant 0$, where we fix $\lambda_{23}=3.14,\lambda_{24}=0.0787$ in the left picture, and $\lambda_{23}=2.4,\lambda_{24}=0.0603$ in the right picture. The colored lines correspond to the viable parameter space of $(\lambda_{13},\lambda_{14})$ satisfying relic density constraint with $m_1=1,2,3$ TeV,  and the respective dashed lines are the upper bound of $\lambda_{14}$ arising from direct detection constraint. }
 \label{fig2}
\end{figure}

In Fig.~\ref{fig2}, we give the results of case (1), where we set $\lambda_{12}=1$ and $\lambda_{13},\lambda_{14} \geqslant 0$. According to Fig.~\ref{fig2}(a), we fix $\lambda_{23}=3.14,\lambda_{24}=0.0787$  while in Fig.~\ref{fig2}(b) we set $\lambda_{23}=2.4,\lambda_{24}=0.0603$. Note that here the values of $\lambda_{24}$ is the upper bound obtained by dark matter direct detection constraint in the absence of $S_1$. The colored lines correspond to the viable parameter space of $(\lambda_{13},\lambda_{14})$ satisfying relic density constraint with $m_1=1,2,3$ TeV,  and the respective dashed lines are the upper bound of $\lambda_{14}$ arising from direct detection constraint. The region of the colored lines below the representative dashed lines correspond to the  parameter space both satisfying relic density and direct detection constraint. Direct detection result puts the most stringent constraint on the parameter space.  Note that $\lambda_{13}$ is related to processes involving $h_2$ while $\lambda_{14}$ is related to processes involving SM Higgs ($h_1$).  According to Fig.~\ref{fig2}, when $\lambda_{13}$ is small, the value of $\lambda_{14}$ is approximately equal to a constant larger than 0.1 under relic density constraint, which means processes related to $h_1$ is dominant over dark matter production while the processes related to $h_2$ are less efficient. With the increase of $\lambda_{13}$, $\lambda_{14}$
decrease sharply as long as $\lambda_{13}$ approximate to 1, where processes related to $h_2$ is so efficient that $\lambda_{13}$ is constrained strictly and $\lambda_{14}$ has a lower bound respectively. 
Notice that direct detection constraint excludes the region that $\lambda_{13}$ is much smaller than 1, which means $h_2$  plays an important role in determining dark matter production within the viable parameter space. For the heavier $m_1$ with $m_1= 2$ TeV and 3 TeV, the upper bound of $\lambda_{13}$ is larger to obtain the correct relic density as the lower bound of $\lambda_{14}$ is larger.
 However, in the case of $m_1= 1$ TeV smaller than $m_{h_2}$, the process of $S_1S_1 \to h_2 h_2$ is highly suppressed and $\lambda_{13}$  should be much larger to meet relic density constraint. Respectively, the lower bound of $\lambda_{14}$ is larger according to Fig.~\ref{fig2}(a). In Fig.~\ref{fig2}(b) with $\lambda_{23}=2.4,\lambda_{24}=0.0603$, we have a similar conclusion, where the viable value of $\lambda_{13}$ is at $\mathcal{O}(1)$ level. However, the lower bound of $\lambda_{14}$ is much larger since the density of $S_2$ is larger in this case and one needs stronger interaction related to $S_1$ to obtain the correct relic density. 
\begin{figure}[htbp]
\centering
 \subfigure[]{\includegraphics[height=6cm,width=6cm]{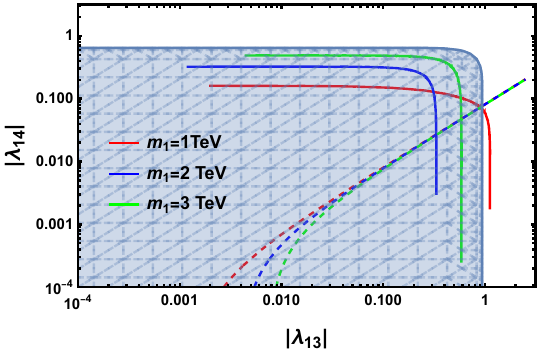}}
\subfigure[]{\includegraphics[height=6cm,width=6cm]{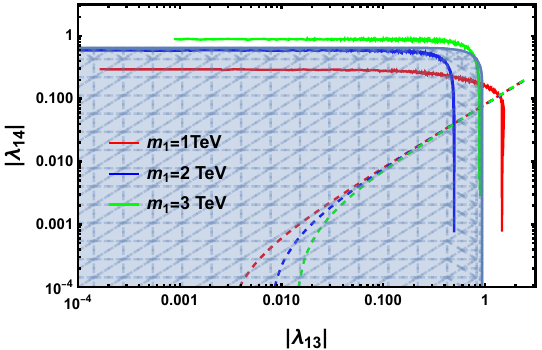}}
\caption{ Results of case (2) with $\lambda_{12}=1,\lambda_{13},\lambda_{14} \leqslant 0$, where we fix $\lambda_{23}=3.14,\lambda_{24}=0.0787$ in the left picture, and $\lambda_{23}=2.4,\lambda_{24}=0.0603$ in the right picture.  The blue-shadowed region corresponds to the parameter space satisfying the copositive criteria of case ii. The colored lines correspond to the viable parameter space of $(|\lambda_{13}|,|\lambda_{14}|)$ satisfying relic density constraint with $m_1=1,2,3$ TeV,  and the respective dashed lines are the upper bound of $|\lambda_{14}|$ arising from direct detection constraint. }
\label{fig4a}
\end{figure}
\begin{figure}[htbp]
\centering
 \subfigure[]{\includegraphics[height=6cm,width=6cm]{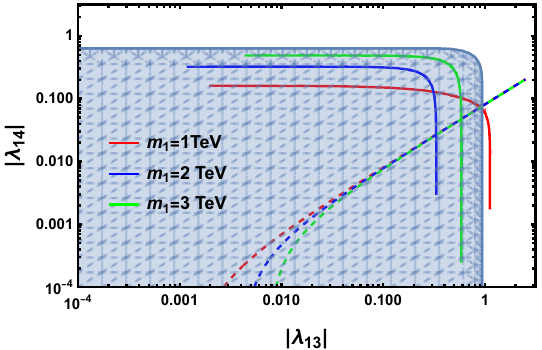}}
\subfigure[]{\includegraphics[height=6cm,width=6cm]{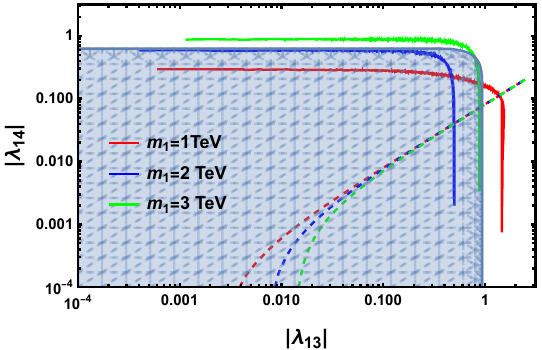}}
\caption{ Results of case (3) with $\lambda_{12}=-1,\lambda_{13},\lambda_{14} \leqslant 0$, where we fix $\lambda_{23}=3.14,\lambda_{24}=0.0787$ in the left picture, and $\lambda_{23}=2.4,\lambda_{24}=0.0603$ in the right picture.  The blue-shadowed region corresponds to the parameter space satisfying the copositive criteria of case iii. The colored lines correspond to the viable parameter space of $(|\lambda_{13}|,|\lambda_{14}|)$ satisfying relic density constraint with $m_1=1,2,3$ TeV,  and the respective dashed lines are the upper bound of $|\lambda_{14}|$ arising from direct detection constraint.}
\label{fig5}
\end{figure}
\begin{figure}[htbp]
\centering
 \subfigure[]{\includegraphics[height=6cm,width=6cm]{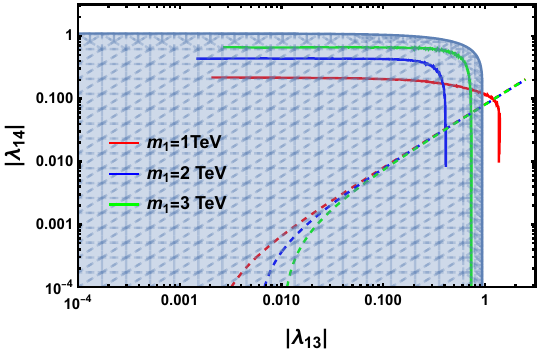}}
\subfigure[]{\includegraphics[height=6cm,width=6cm]{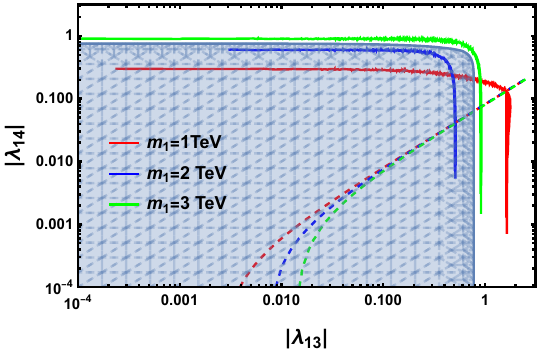}}
\caption{ Results of case (4) with $\lambda_{12}=1,\lambda_{13},\lambda_{14},\lambda_{23},\lambda_{24}\leqslant 0$, where we fix $\lambda_{23}=-0.9,\lambda_{24}=-0.023$ in the left picture, and $\lambda_{23}=-0,75,\lambda_{24}=-0.194$ in the right picture.  The blue-shadowed region corresponds to the parameter space satisfying the copositive criteria of case iii. The colored lines correspond to the viable parameter space of $(|\lambda_{13}|,|\lambda_{14}|)$ satisfying relic density constraint with $m_1=1,2,3$ TeV,  and the respective dashed lines are the upper bound of $|\lambda_{14}|$ arising from direct detection constraint.}
\label{fig6}
\end{figure}

 We give the results of case (2) in Fig.~\ref{fig4a} with $\lambda_{12}=1,\lambda_{13},\lambda_{14} \leqslant 0$, where we fix $\lambda_{23}=3.14,\lambda_{24}=0.0787$ in the left picture, and $\lambda_{23}=2.4,\lambda_{24}=0.0603$ in the right picture.   
 The colored lines correspond to the viable parameter space of $(|\lambda_{13}|,|\lambda_{14}|)$ satisfying relic density constraint with $m_1=1,2,3$ TeV,  and the respective dashed lines are the upper bound of $|\lambda_{14}|$ arising from direct detection constraint.  The blue-shadowed region corresponds to the parameter space satisfying copositive criteria. Since the cross-section of $S_iS_i \to h_i h_i(i=1,2)$ is determined by Higgs-mediated processes, the quartic term $|S_i|^2|h_i|^2$ and the t-channel processes exchange by dark matter, where the sign of the couplings $\lambda_{13}$ and $\lambda_{14}$  can make difference in the interference terms of these diagrams, therefore the viable parameter space will be different.   According to Fig.~\ref{fig4a}(a), the smaller $|\lambda_{13}|$ is excluded for dark matter density being over-abundant due to the interference effect. Similarly,
  the viable $|\lambda_{13}|$ value is constrained within a narrow region while $|\lambda_{14}$ value is more flexible but sharply decreases with the increase of $|\lambda_{13}|$ by direct detection constraint as the results of case i, where $h_2$ plays a dominant role in determining dark matter production. On the other hand,  for the heavier $m_1$, the upper bound of $|\lambda_{13}|$ is much smaller to obtain the correct relic density compared with the results of case i in Fig.~\ref{fig2}(a) due to the interference effect. What's more, copositive criteria 
 in case ii can constrain the parameter space strictly and for $m_1=1$ TeV, the allowed $(|\lambda_{13}|,|\lambda_{14}|)$ is almost excluded as we can see in  Fig.~\ref{fig4a}(a).  For   Fig.~\ref{fig4a}(b) with $\lambda_{23}=2.4,\lambda_{24}=0.0603$, the copositive criterial naturally excludes small $|\lambda_{13}|$ from the point of theoretical constraint in the case of $m_1=3 $ TeV. Moreover, for $m_1=1$ TeV, the allowed parameter space of $(|\lambda_{13}|,|\lambda_{14}|)$ are completely excluded by the copositive criterial.
 
 In Fig.~\ref{fig5} (a)and Fig.~\ref{fig5}(b),  we give the results of case (3) with $\lambda_{12}=-1,\lambda_{13},\lambda_{14} \leqslant 0$, where we fix $\lambda_{23}=3.14,\lambda_{24}=0.0787$ in the left picture, and $\lambda_{23}=2.4,\lambda_{24}=0.0603$ in the right picture.   The difference between case (3) and case (2) is $\lambda_{12}$ set to be negative, the combined contribution of the Higgs-mediated processes and the quartic term $S_1^2S_2^2$ will determine the cross-section of $S_1S_1 \to S_2S_2$, and the sign of $\lambda_{12}$ will affect the interference terms.  However, such interactions between $S_1$ and $S_2$ just adjust he relative relic density of $S_1$ and $S_2$,  and make little difference in the viable parameter space of $\lambda_{13}$ and $\lambda_{14}$. Therefore, we almost have similar results with case (2) as we can see in Fig.~\ref{fig5} (a)and Fig.~\ref{fig5}(b).
 
 We give the results of case (4) in Fig.~\ref{fig6} with $\lambda_{12}=1,\lambda_{13},\lambda_{14} ,\lambda_{23},\lambda_{24} \leqslant 0$. According to the copositive criteria of Case VIII, the value of $\lambda_{23}$ and $\lambda_{24}$ are also constrained. Within the chosen parameter space, one can obtain $|\lambda_{23}|\leqslant 0.94$ and $|\lambda_{24}|\leqslant 0.64$ with Eq.~\ref{eq21}. In Fig.~\ref{fig6}(a), we fix $\lambda_{23}=-0.9$ and $\lambda_{24}=-0.023$ and the allowed values of $|\lambda_{13}|$ as well as $|\lambda_{14}|$ are completely excluded by direct detection constraint and copositive criteria within the chosen parameter space in the case of $m_1= 1$ TeV. On the other hand, for $m_1=3$ TeV, the allowed value of $|\lambda_{14}|$ is more flexible due to the interference effect. We set $\lambda_{23}=-0.75$ and $\lambda_{24}=-0.194$ in Fig.~\ref{fig6}(b), where copositive criterial not just excludes the  parameter space of $(|\lambda_{13}|,|\lambda_{14}|)$ in the case of $m_1=1$ TeV
but also excludes the case of $m_1=3$ TeV regardless of direct detection constraint.
   
   In conclusion, according to Fig.~\ref{fig2} to Fig.~\ref{fig6}, we show the effect of copositive criteria on the parameter space.  Different choices of the signs of the couplings will contribute to different copositive conditions, which can induce different viable parameter spaces due to the interference effect. We consider 4 different cases and randomly scan the chosen parameter space under dark matter relic density and direct detection constraints.  Direct detection constraint puts the most stringent limit on the parameter space, which demands new Higgs $h_2$ plays a dominant role in determining dark matter relic density. On the other hand, copositive criterial can also constrain the allowed parameters space that the cases of $m_1=1$ TeV are excluded for different copositive conditions when some of the couplings are negative. Particularly, when we consider $\lambda_{12}=1,\lambda_{13},\lambda_{14} \leqslant 0, \lambda_{23}=-0.75$ and $\lambda_{24}=-0.194$, copositive criteria is more stringent on the viable parameters and the parameter space of $m_1=3$ TeV is completely excluded regardless of direct detection constraint. The high values of the couplings $\sim 1$  can imply a  low energy validity for the model with two stable scalars, which will set limitations for the model at high energies. In the appendix, we perform a random scan and the viable parameter spaces with  four different cases  satisfying copositive criterial, relic density and direct detection constraint are shown. According to the results, the allowed values for the couplings can be at $\mathcal{O}(0.1)$ level, which are okay with high scale validity of the model.

\section{Summary and Outlook}\label{sec:sum}
Weakly interacting massive particles as dark matter are facing serious challenges that no signals have been found in the direct detection experiments so far, and one solution to such a problem is multi-component dark matter models, which include two or more dark matter species. On the other hand, the multi-component dark matter should be constrained by theoretical constraints and experiment constraints. Particularly, for the extended scalar dark matter models. the vacuum stability demand the scalar potential has to be bounded from below, which can put stringent constraints on the parameter space. What's more, the copositive criteria allow the derivation analytic necessary and sufficient vacuum stability conditions for the couplings, and different copositivity criteria will contribute to different viable parameter spaces depending on the choice of the signs of the couplings.

 We consider a two-component scalar dark matter model in this work. To obtain a stable vacuum, the $4\times 4$ matrix of the quadratic couplings should satisfy copositivity criteria. Provided that direct detection results put stringent constraints on the parameter space, discussion about the related parameters can indeed be simplified greatly and we can just focus on 4 different cases related to copositive criteria depending on the signs of the quadratic couplings in the model. The signs of these parameters will not only demand different copositivity criteria but can also influence the results of the cross-section of the $2 \to 2$ processes and induce different viable parameter space due to the interference effect.  We randomly scan the chosen parameter space with dark matter relic density and direct detection constraints. For the 4 different cases, the allowed values of $(|\lambda_{13}|,|\lambda_{14}|)$ are different but all demand the new Higgs $h_2$ play a dominant role in determining dark matter relic density within the viable parameter space. In addition, we perform a random scan in the appendix under dark matter relic density and direct detection constraint, there are a few points "survived" when copositive criterial is considered, and for different criterials, the allowed values for the couplings are different according to the results. In other word, the copositive criteria can also put stringent constraints on the parameter space as long as some quartic coupling is negative. 
 
  As we can see in this work,  different choices of the signs of couplings can indeed contribute to different parameter spaces with copositive criteria. The model in this work is simple but the discussion about copositive criterial is complex for the different permutations of the indexes.
 For a more complex model, copositivity allows us to find analytic necessary and sufficient vacuum stability conditions so that estimate the parameter space theoretically. 
\begin{acknowledgments}
\noindent
Hao Sun is supported by the National Natural Science Foundation of China (Grant No. 12075043, No. 12147205). XinXin Qi is supported by the National Natural Science Foundation of China (Grant
No.12447162).
\end{acknowledgments}
\appendix
\section{Appendix}
\label{appA}
\subsection{Perturbativity}
To ensure the perturbative model, the contribution
from loop correction should be smaller than the tree level
values, and such constraints can be ensured with 
\begin{eqnarray}
|\lambda_{13}|<4\pi,|\lambda_{23}|<4\pi,|\lambda_{12}|<4\pi,|\lambda_{14}|<4\pi,|\lambda_{24}|<4\pi.
\end{eqnarray}
\subsection{ Perturbativity unitarity}
The unitarity conditions come from the tree-level scalar-scalar scattering matrix which is
dominated by the quartic contact interaction. The s-wave scattering amplitudes should
lie under the perturbative unitarity limit, given the requirement the eigenvalues of the
S-matrix $\mathcal{M}$ must be less than the unitarity bound given by $|\mathrm{Re}\mathcal{M}| < \frac{1}{2}$.

\subsection{Renormalization Group Equations of the quartic couplings }
In this part, we give the renormalization Group Equations (RGEs) of the model with SARAH\cite{Staub:2015kfa} at one-loop level, and the beta functions of the quartic couplings are given by:
\begin{align}
\beta_{\lambda_{24}} & =  
+2 \lambda_{12} \lambda_{14}-\frac{9}{10} g_{1}^{2} \lambda_{24} -\frac{9}{2} g_{2}^{2} \lambda_{24} +8 \lambda_{22} \lambda_{24} +4 \lambda_{24}^{2} +2 \lambda_{23} \lambda_{34} +12 \lambda_{24} \lambda_{44} +6 \lambda_{24} y_t^2, \\ 
\beta_{\lambda_{14}} & =  
-\frac{9}{10} g_{1}^{2} \lambda_{14} -\frac{9}{2} g_{2}^{2}\lambda_{14} +8 \lambda_{11} \lambda_{14} +4 \lambda_{14}^{2} +2 \lambda_{12} \lambda_{24} +2 \lambda_{13} \lambda_{34} +12 \lambda_{14} \lambda_{44}  +6 \lambda_{14}y_t^2, \\ 
\beta_{\lambda_{12}} & =  
2 \lambda_{13} \lambda_{23} + 4 \lambda_{12}^{2}  + 4 \lambda_{14} \lambda_{24}  + 8 \lambda_{11} \lambda_{12} + 8 \lambda_{12} \lambda_{22}, \\ 
\beta_{\lambda_{23}} & =  
2 \lambda_{12} \lambda_{13}  + 4 \Big(2 \lambda_{22}\lambda_{23}  + 2 \lambda_{23} \lambda_{33}  + \lambda_{24}\lambda_{34}  + \lambda_{23}^{2}\Big),\\ 
\beta_{\lambda_{13}} & =  
2 \lambda_{12}\lambda_{23} + 4\lambda_{13}^{2}  + 4 \lambda_{14} \lambda_{34}  + 8 \lambda_{11}\lambda_{13} + 8 \lambda_{13}\lambda_{33}.
\end{align}
where $y_t$ is the Top Yukawa coupling, $g_1$ and $g_2$ are the gauge couplings of $U(1)$ and $SU(2)$.
\subsection{Cross section of $S_iS_i \to h_ih_i$ and $S_iS_i \to h_1h_2 (i=1,2)$ }\label{csa}
We give the cross section of $S_iS_i \to h_ih_i$ and $S_iS_i \to h_1h_2$ $(i=1,2)$ in the part.  We calculate the cross section with Calchep \cite{Belyaev:2012qa}, and the expressions in the limit of $\sin\theta\to 0$ are given as follows.\\
\begin{align}
\sigma_{S_1S_1 \to h_1h_1}&=\frac{\lambda_{14}^2}{8 \pi s (s-4 m_1^2)} (\sqrt{(s-4 m_{h_1}^2) (s-4 m_1^2)} (\frac{8 \lambda_{14}^2 v_0^4}{m_{h_1}^4-4 m_{h_1}^2 m_1^2+m_1^2 s}+\frac{(2 m_{h_1}^2+s)^2}{(m_{h_1}^2-s)^2})\notag \\
 + &\frac{8 \lambda_{14} v_0^2 (2 \lambda_{14} v_0^2 (m_{h_1}^2-s)-4 m_{h_1}^4+s^2)}{2m_{h_1}^4-3 m_{h_1}^2 s+s^2} \log (\frac{2 m_{h_1}^2+s (\sqrt{\frac{(s-4 m_{h_1}^2) (s-4 m_1^2)}{s^2}}-1)}{2 m_{h_1}^2-s (\sqrt{\frac{(s-4 m_1^2) (s-4 m_{h_1}^2)}{s^2}}+1)})),
\end{align}
\begin{align}
\sigma_{S_1S_1 \to h_2 h_2} &= \frac{\lambda_{13}^2}{8 \pi s (s-4 m_1^2)} (\sqrt{(s-4 m_{h_2}^2) (s-4 m_1^2)} (\frac{8 \lambda_{13}^2 v_1^4}{m_{h_2}^4-4 m_{h_2}^2 m_1^2+m_1^2 s}+\frac{(2 m_{h_2}^2+s)^2}{(m_{h_2}^2-s)^2})\notag \\
 + &\frac{8 \lambda_{13} v_1^2 (2 \lambda_{13} v_1^2 (m_{h_2}^2-s)-4 m_{h_2}^4+s^2)}{2m_{h_2}^4-3 m_{h_2}^2 s+s^2} \log (\frac{2 m_{h_2}^2+s (\sqrt{\frac{(s-4 m_{h_2}^2) (s-4 m_1^2)}{s^2}}-1)}{2 m_{h_2}^2-s (\sqrt{\frac{(s-4 m_1^2) (s-4 m_{h_2}^2)}{s^2}}+1)})),
\end{align}
\begin{align}
\sigma_{S_2S_2 \to h_1h_1} &=\frac{\lambda_{24}^2}{8 \pi s (s-4 m_2^2)} (\sqrt{(s-4 m_{h_1}^2) (s-4 m_2^2)} (\frac{8 \lambda_{24}^2 v_0^4}{m_{h_1}^4-4 m_{h_1}^2 m_2^2+m_2^2 s}+\frac{(2 m_{h_1}^2+s)^2}{(m_{h_1}^2-s)^2})\notag \\
 + &\frac{8 \lambda_{24} v_0^2 (2 \lambda_{24} v_0^2 (m_{h_1}^2-s)-4 m_{h_1}^4+s^2)}{2m_{h_1}^4-3 m_{h_1}^2 s+s^2} \log (\frac{2 m_{h_1}^2+s (\sqrt{\frac{(s-4 m_{h_1}^2) (s-4 m_2^2)}{s^2}}-1)}{2 m_{h_1}^2-s (\sqrt{\frac{(s-4 m_2^2) (s-4 m_{h_1}^2)}{s^2}}+1)})),
\end{align}
\begin{align}
\sigma_{S_2S_2 \to h_2 h_2} &=\frac{\lambda_{23}^2}{8 \pi s (s-4 m_2^2)} (\sqrt{(s-4 m_{h_2}^2) (s-4 m_2^2)} (\frac{8 \lambda_{23}^2 v_1^4}{m_{h_2}^4-4 m_{h_2}^2 m_2^2+m_2^2 s}+\frac{(2 m_{h_2}^2+s)^2}{(m_{h_2}^2-s)^2})\notag \\
 + &\frac{8 \lambda_{23} v_1^2 (2 \lambda_{23} v_1^2 (m_{h_2}^2-s)-4 m_{h_2}^4+s^2)}{2m_{h_2}^4-3 m_{h_2}^2 s+s^2} \log (\frac{2 m_{h_2}^2+s (\sqrt{\frac{(s-4 m_{h_2}^2) (s-4 m_2^2)}{s^2}}-1)}{2 m_{h_2}^2-s (\sqrt{\frac{(s-4 m_2^2) (s-4 m_{h_2}^2)}{s^2}}+1)})),
\end{align}
\begin{align}
\sigma_{S_1S_1 \to h_1 h_2} &=\frac{4 \lambda_{13}^2 \lambda_{14}^2 v_0^2 v_1^2}{\pi s (s-4 m_1^2)} \frac{\log (\frac{m_{h_1}^2+s (\sqrt{\frac{(s-4 m_1^2) (m_{h_1}^4-2 m_{h_1}^2 (m_{h_2}^2+s)+(m_{h_2}^2-s)^2)}{s^3}}-1)+m_{h_2}^2}{m_{h_1}^2-s (\sqrt{\frac{(s-4 m_1^2) (m_{h_1}^4-2 m_{h_1}^2 (m_{h_2}^2+s)+(m_{h_2}^2-s)^2)}{s^3}}+1)+m_{h_2}^2})}{m_{h_1}^2+m_{h_2}^2-s} \notag\\
+&\frac{s \sqrt{\frac{(s-4m_1^2) (m_{h_1}^4-2 m_{h_1}^2 (m_{h_2}^2+s)+(m_{h_2}^2-s)^2)}{s}}}{2 (-2m_1^2 s (m_{h_1}^2+m_{h_2}^2)+m_1^2 (m_{h_1}^2-m_{h_2}^2)^2+m_{h_1}^2 m_{h_2}^2 s+m_1^2 s^2)},
\end{align}
\begin{align}
\sigma_{S_2S_2 \to h_1 h_2} &=\frac{4 \lambda_{23}^2 \lambda_{24}^2 v_0^2 v_1^2}{\pi s (s-4 m_2^2)} \frac{\log (\frac{m_{h_1}^2+s (\sqrt{\frac{(s-4 m_2^2) (m_{h_1}^4-2 m_{h_1}^2 (m_{h_2}^2+s)+(m_{h_2}^2-s)^2)}{s^3}}-1)+m_{h_2}^2}{m_{h_1}^2-s (\sqrt{\frac{(s-4 m_2^2) (m_{h_1}^4-2 m_{h_1}^2 (m_{h_2}^2+s)+(m_{h_2}^2-s)^2)}{s^3}}+1)+m_{h_2}^2})}{m_{h_1}^2+m_{h_2}^2-s} \notag\\
+&\frac{s \sqrt{\frac{(s-4m_2^2) (m_{h_1}^4-2 m_{h_1}^2 (m_{h_2}^2+s)+(m_{h_2}^2-s)^2)}{s}}}{2 (-2m_2^2 s (m_{h_1}^2+m_{h_2}^2)+m_2^2 (m_{h_1}^2-m_{h_2}^2)^2+m_{h_1}^2 m_{h_2}^2 s+m_2^2 s^2)}.
\end{align}
According to the expressions of the cross-section, in the limit of $\sin\theta \to 0$, the signs of $\lambda_{13}$,$\lambda_{23}$,$\lambda_{14}$ and $\lambda_{24}$ can make difference on the terms including the odd orders of these couplings so that contribute to the cross-section of $S_iS_i \to h_ih_i (i=1,2)$. Note that for the non-zero $\sin\theta$, the sign of $\lambda_{13}$,$\lambda_{23}$,$\lambda_{14}$ and $\lambda_{24}$ can make also difference in the cross section by the combined contributions of these couplings such as $\lambda_{13}\lambda_{24}$.
\begin{figure}[htbp]
\centering
 \subfigure[]{\includegraphics[height=6cm,width=6cm]{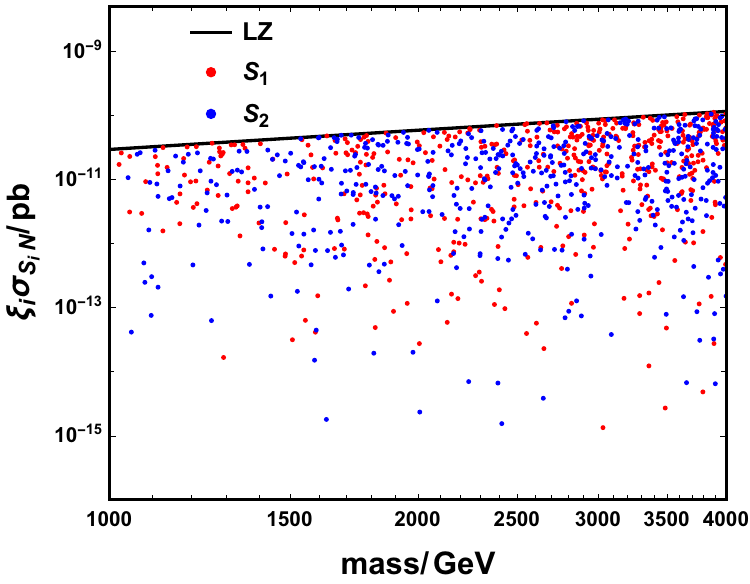}}
 \subfigure[]{\includegraphics[height=6.5cm,width=6cm]{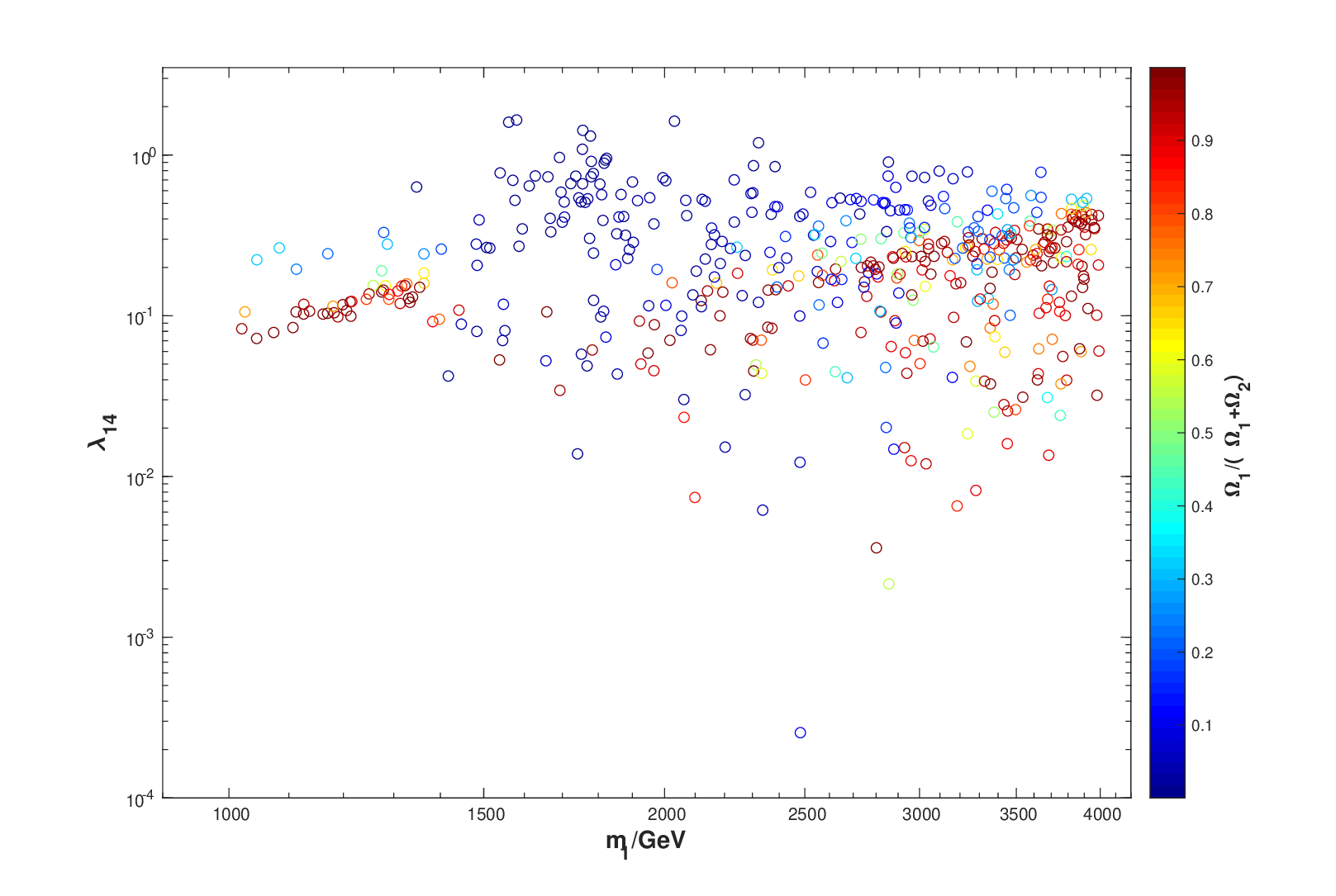}}
\subfigure[]{\includegraphics[height=6cm,width=6cm]{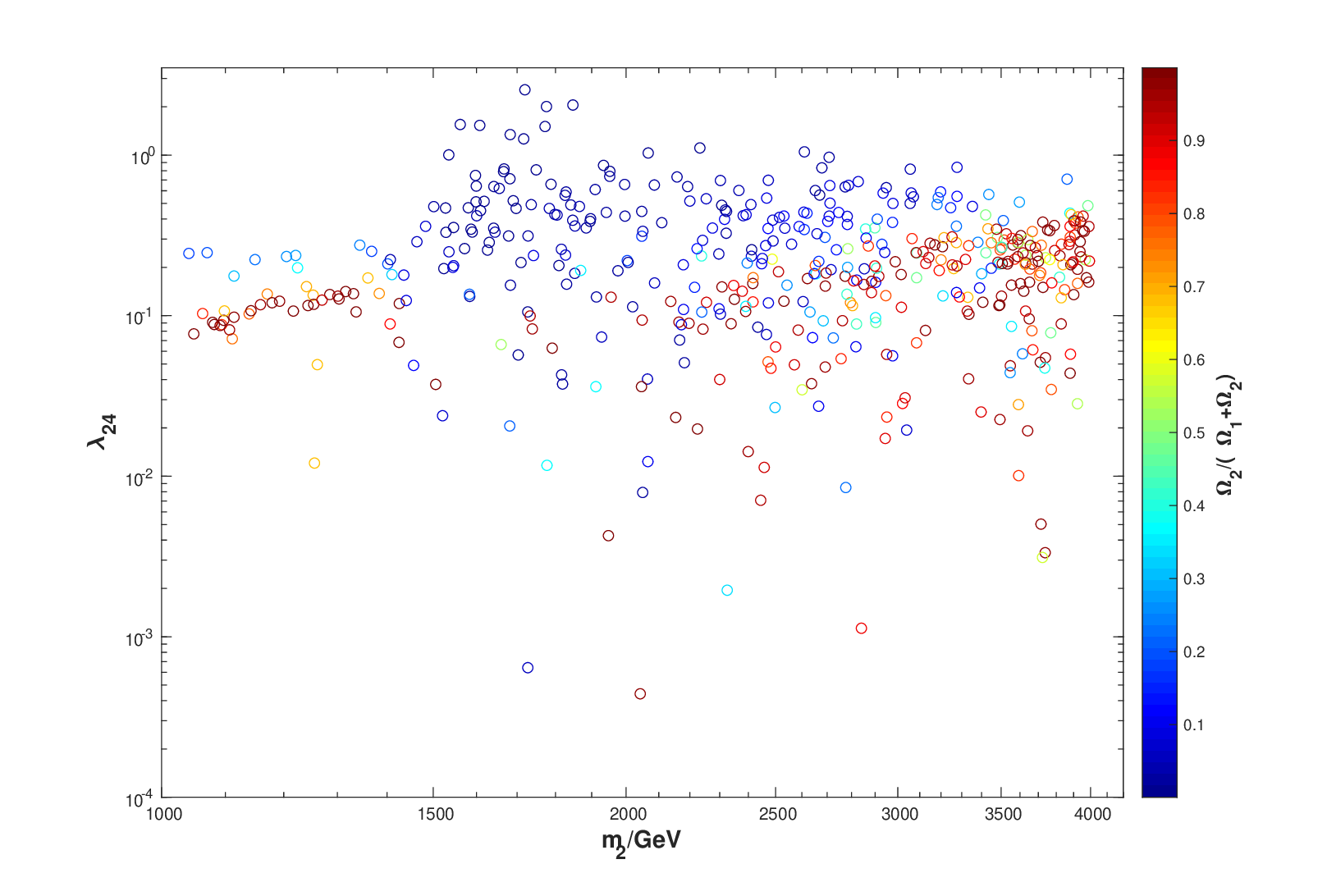}}
 \subfigure[]{\includegraphics[height=6cm,width=6cm]{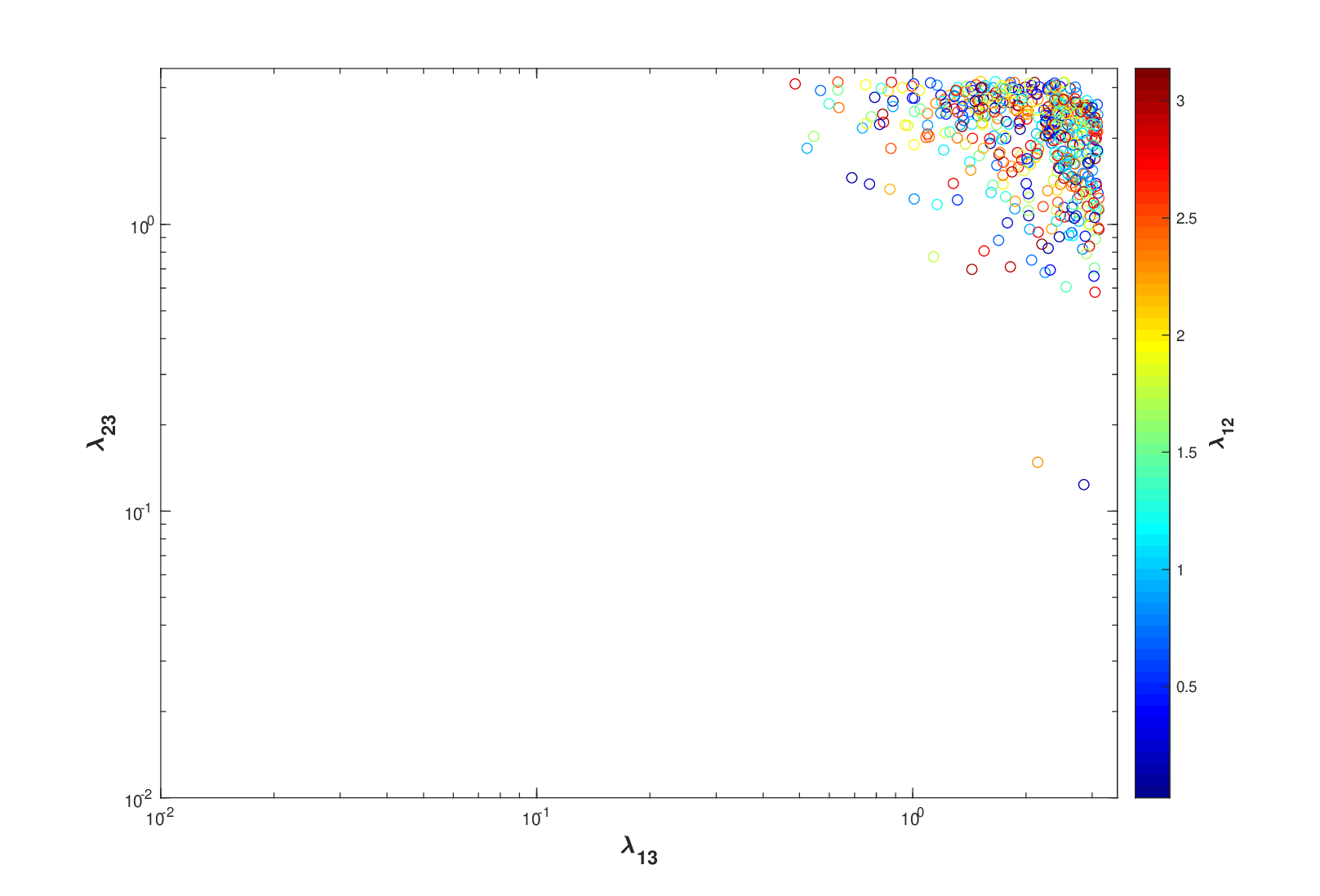}}
\caption{Results of viable parameter space  satisfying the relic density and the direct detection constraint for case (1). The picture (a) shows the  allowed values for dark matter mass $S_1$(red points) and $S_2$ (blue points), and the black line represents the upper bound according to the latest LZ results. The picture (b) shows the viable parameter space for $m_1-\lambda_{14}$, where bubbles with different colors correspond to the fraction of $S_1$ defined by $\Omega_1/(\Omega_1+\Omega_2)$. The picture (c) shows the viable parameter space for $m_2-\lambda_{24}$, where bubbles with different colors correspond to the fraction of $S_2$ defined by $\Omega_2/(\Omega_1+\Omega_2)$. The picture (d) shows the allowed value for $\lambda_{13}-\lambda_{23}$, where bubbles with different colors represent the viable value for $\lambda_{12}$. }
\label{figs1co}
\end{figure}

\begin{figure}[htbp]
\centering
\subfigure[]{\includegraphics[height=6cm,width=6cm]{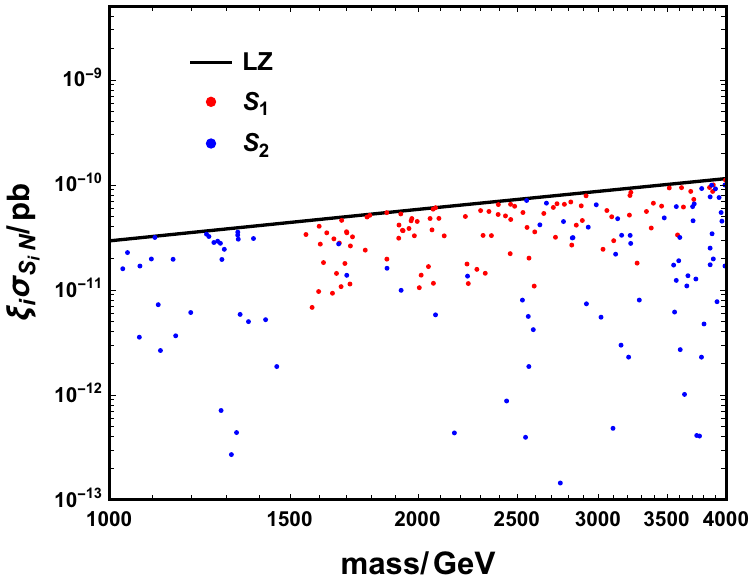}}
 \subfigure[]{\includegraphics[height=6.5cm,width=6cm]{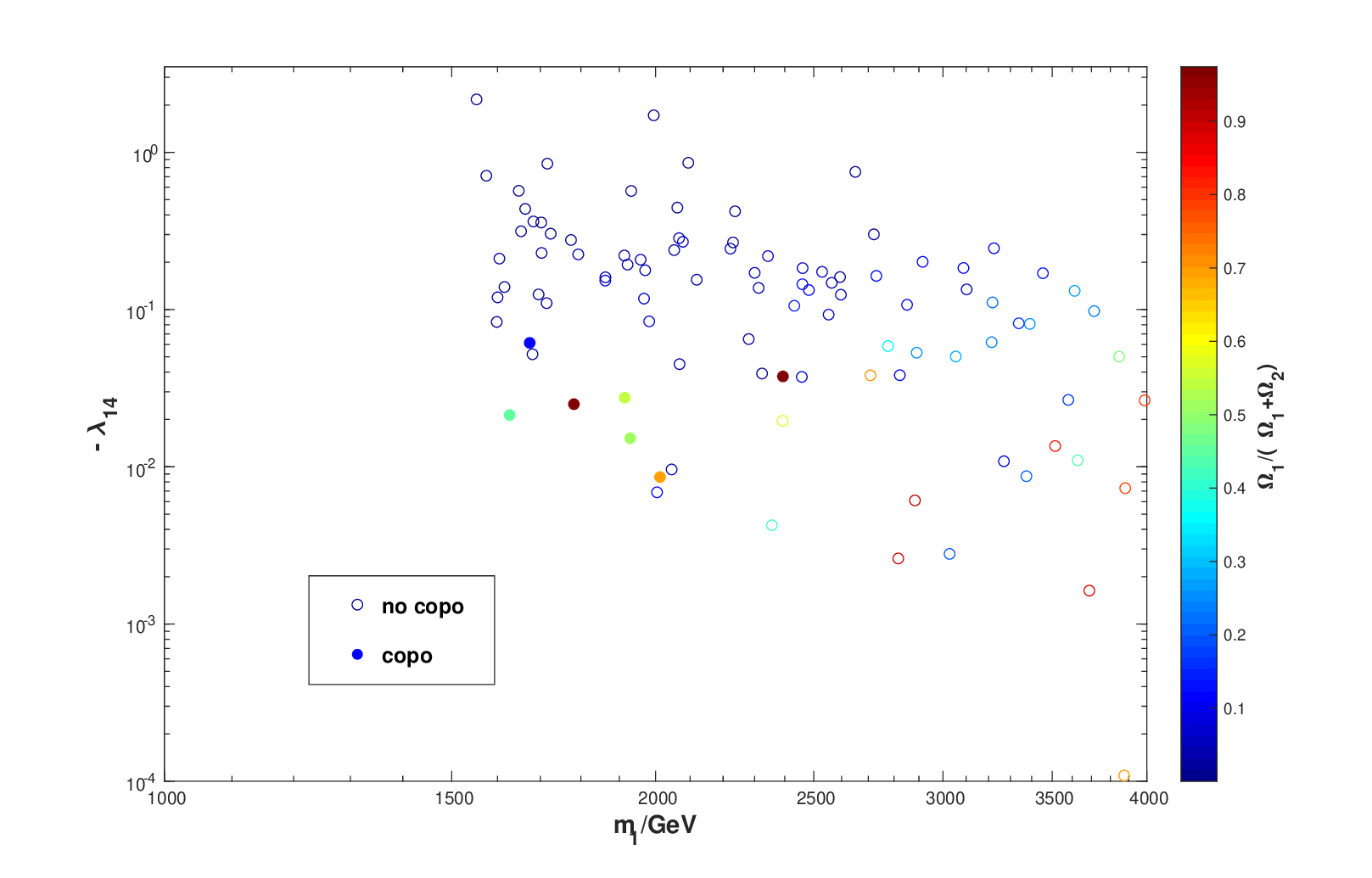}}
\subfigure[]{\includegraphics[height=6cm,width=6cm]{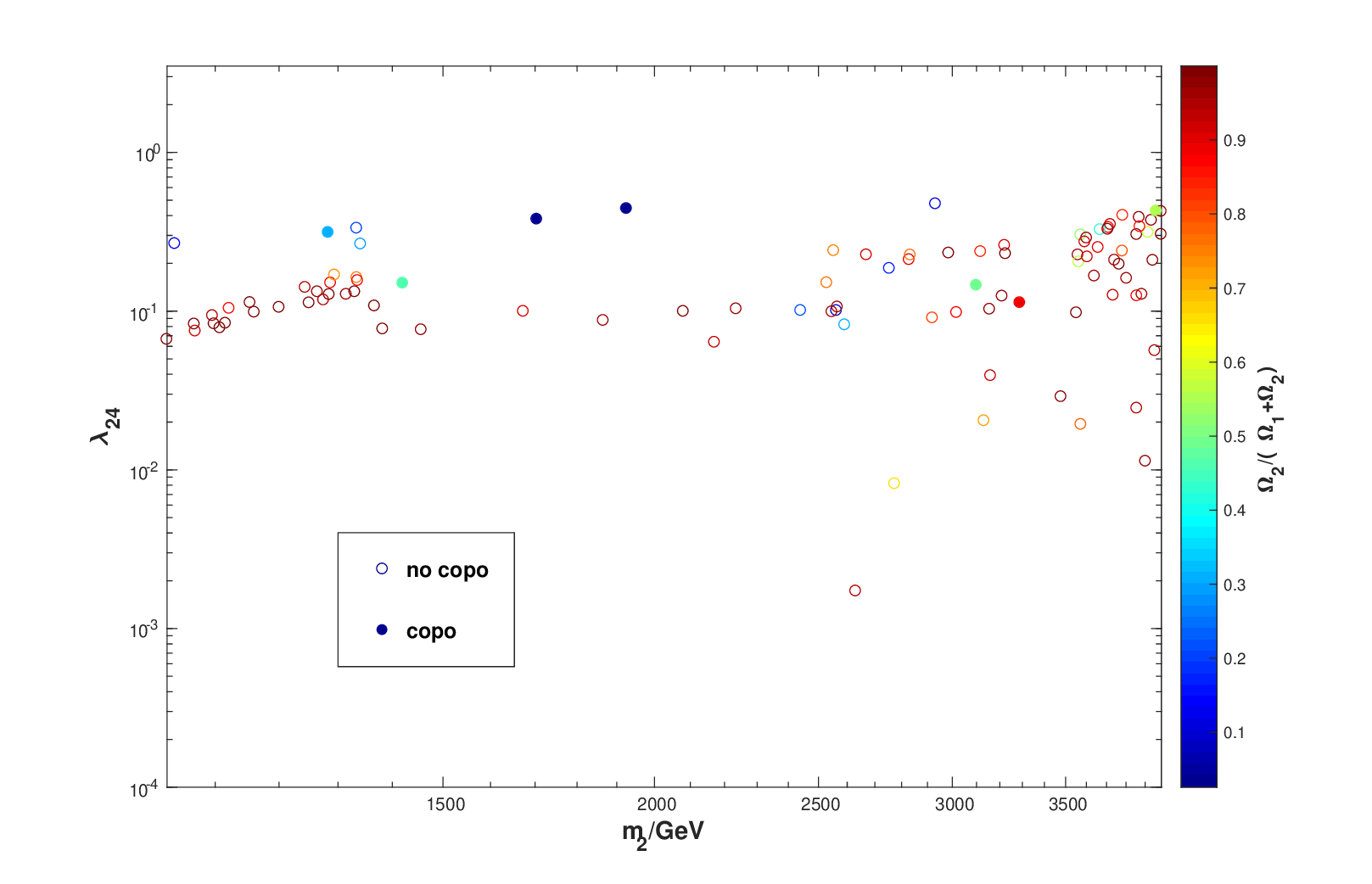}}
 \subfigure[]{\includegraphics[height=6cm,width=6cm]{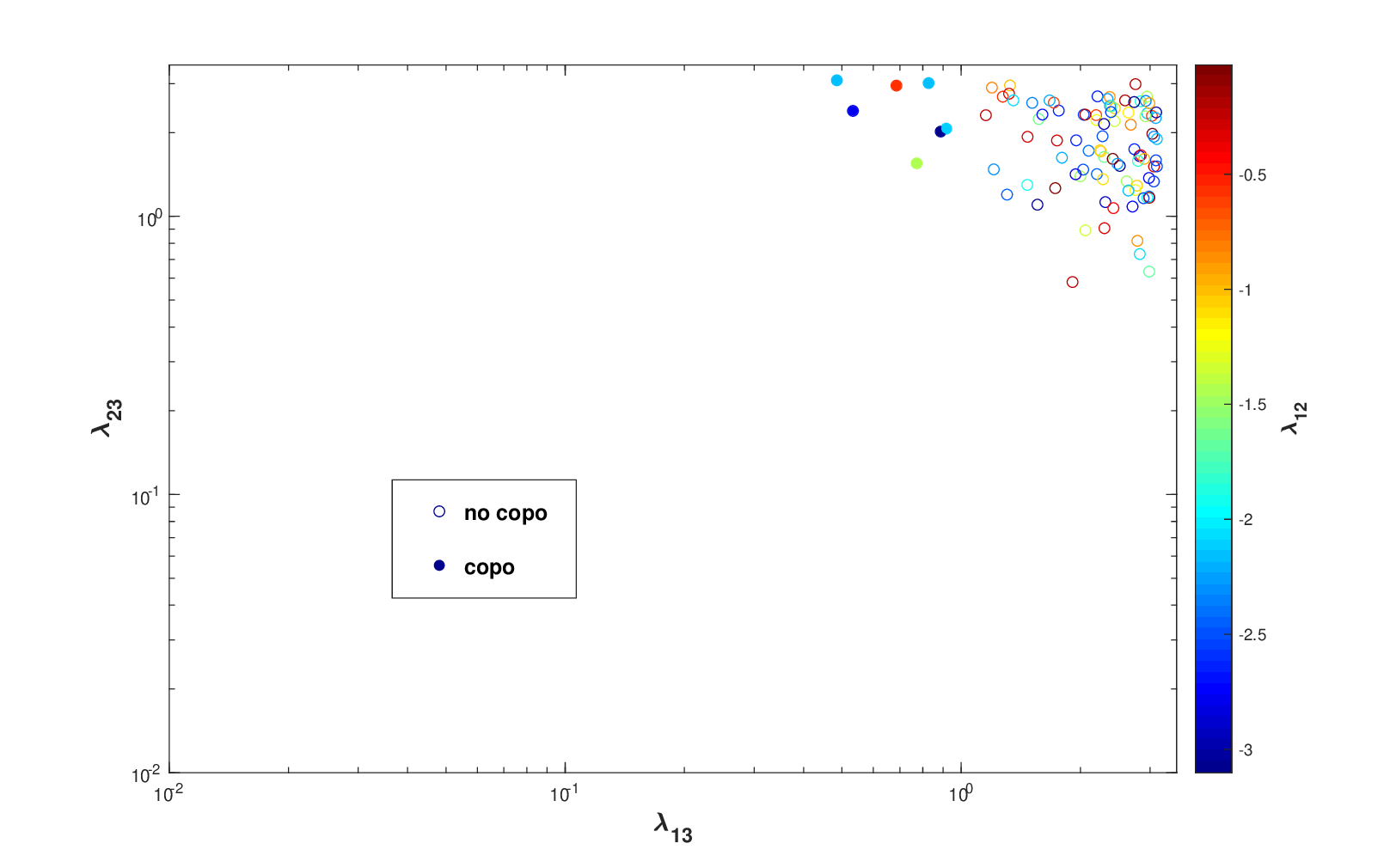}}
\caption{Results of viable parameter space  satisfying the relic density and the direct detection constraint for case (2). In (b),(c),(d), the colored points satisfying copositive criterial while the colored bubbles are not, which are labeled by "copo" and "no copo".}
\label{figs4co}
\end{figure}
\begin{figure}[htbp]
\centering
\subfigure[]{\includegraphics[height=6cm,width=6cm]{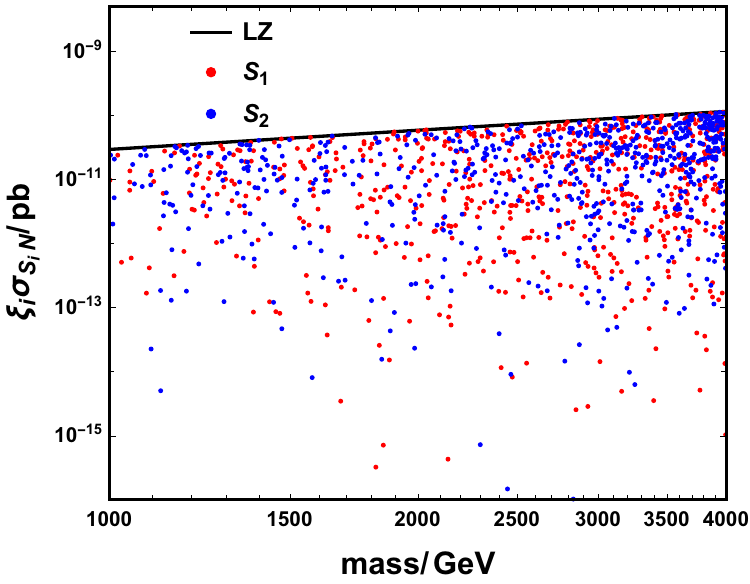}}
 \subfigure[]{\includegraphics[height=6.5cm,width=6cm]{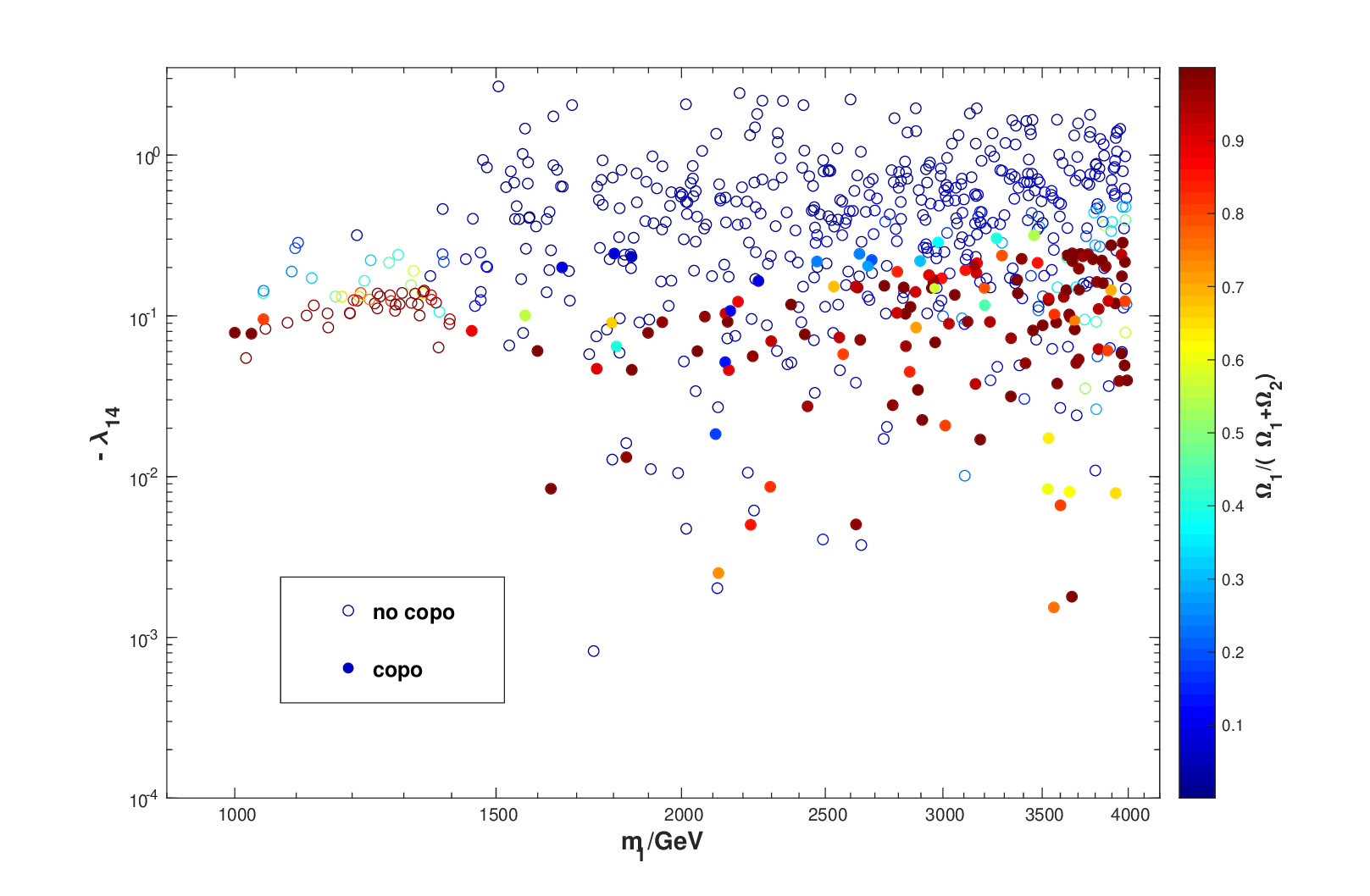}}
\subfigure[]{\includegraphics[height=6cm,width=6cm]{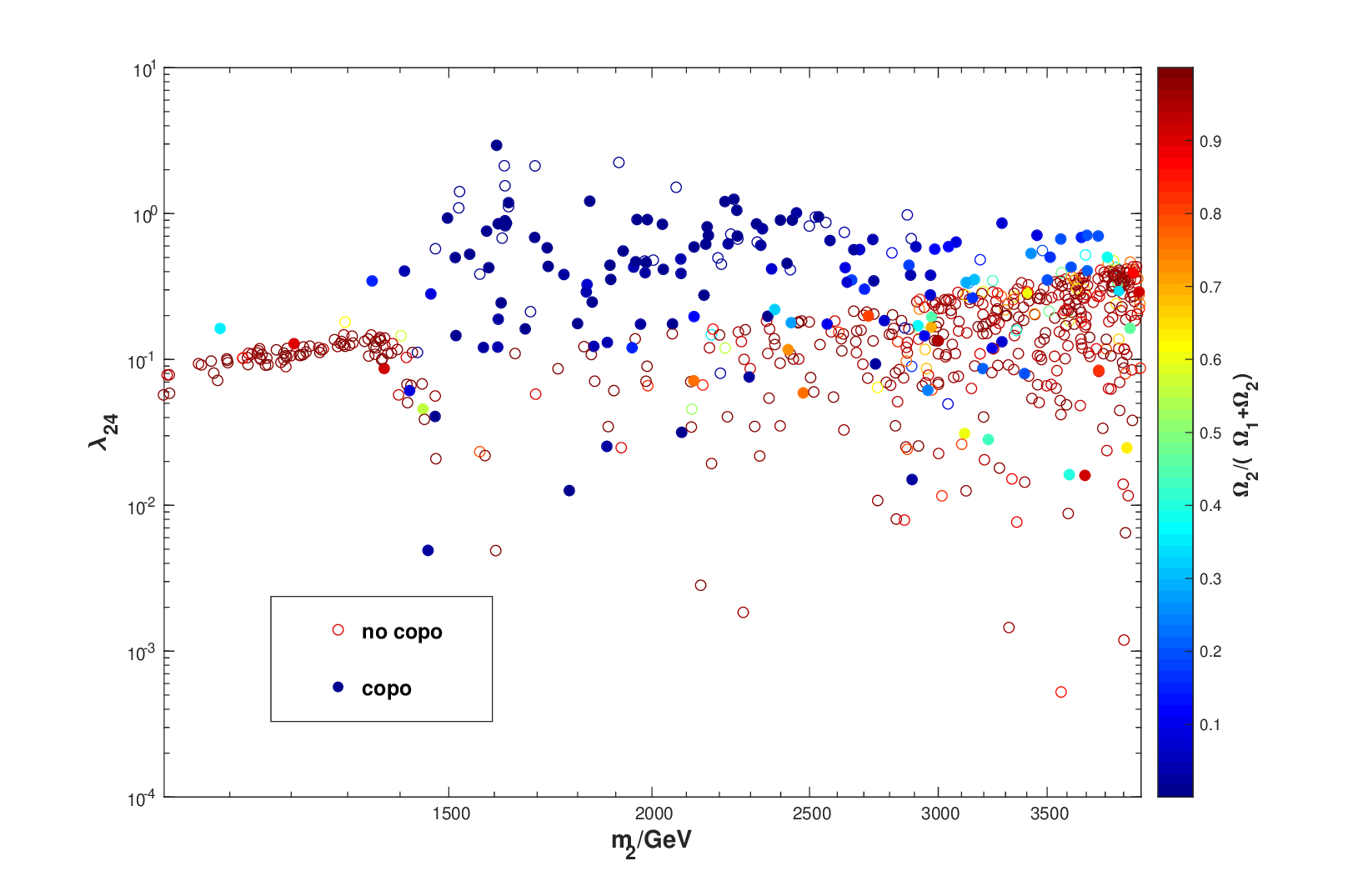}}
 \subfigure[]{\includegraphics[height=6cm,width=6cm]{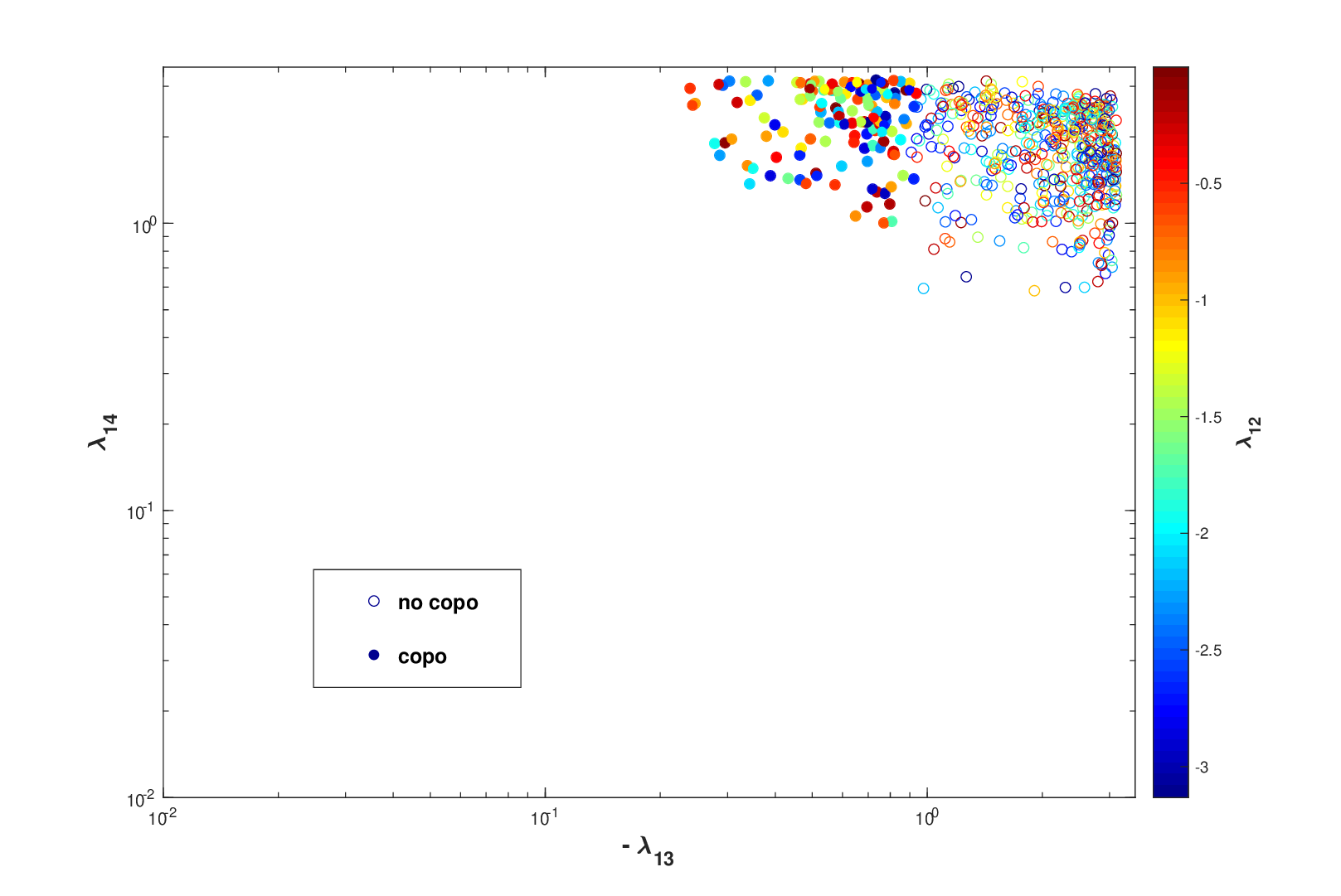}}
\caption{Results of viable parameter space  satisfying the relic density and the direct detection constraint for case (3). In (b),(c),(d), the colored points satisfying copositive criterial while the colored bubbles are not, which are labeled by "copo" and "no copo".}
\label{figs6co}
\end{figure}
\begin{figure}[htbp]
\centering
\subfigure[]{\includegraphics[height=6cm,width=6cm]{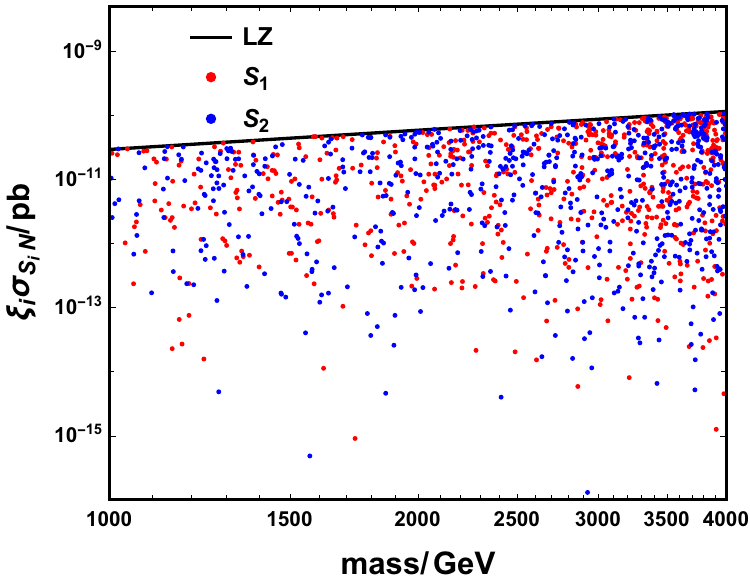}}
 \subfigure[]{\includegraphics[height=6.5cm,width=6cm]{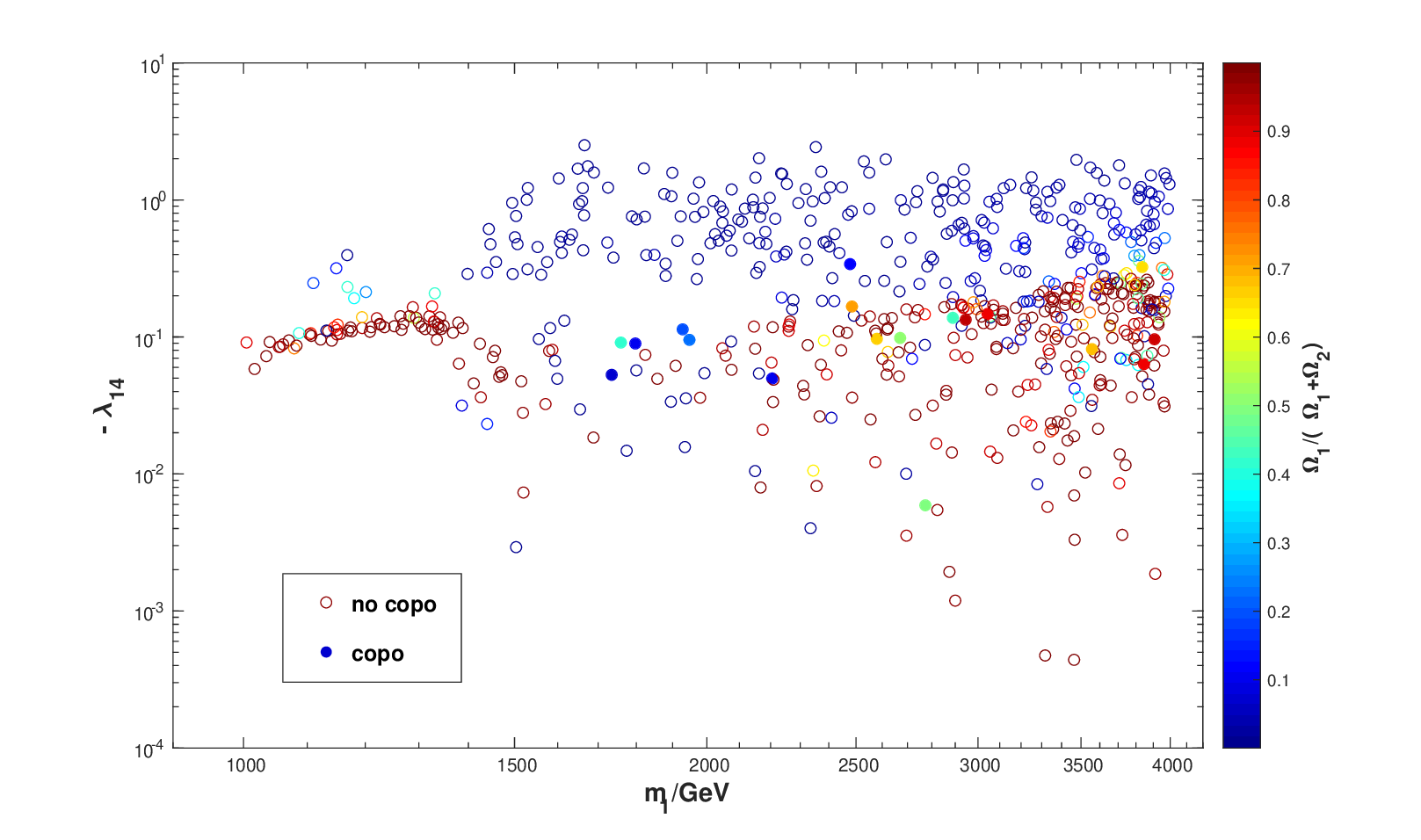}}
\subfigure[]{\includegraphics[height=6cm,width=6cm]{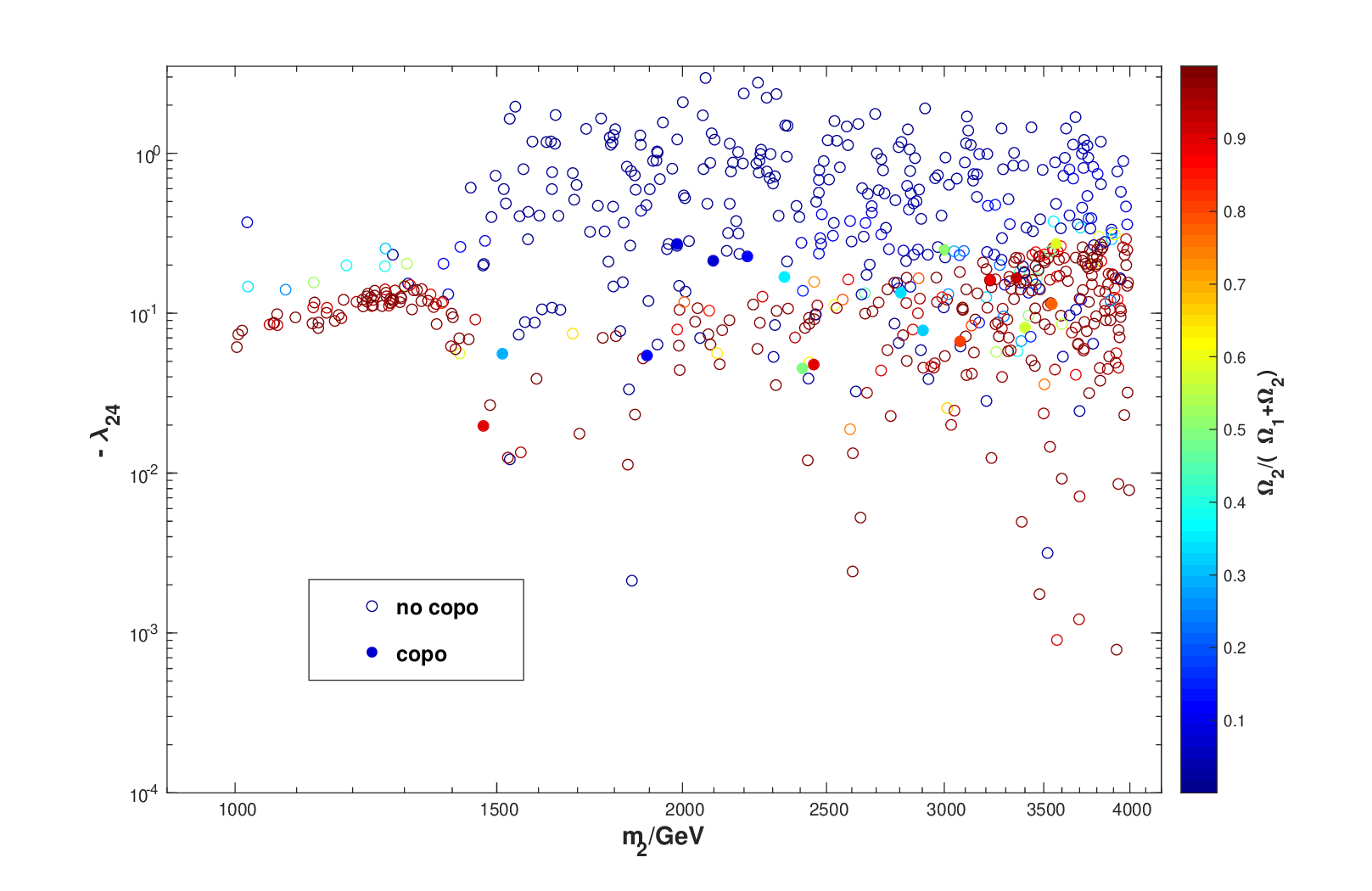}}
 \subfigure[]{\includegraphics[height=6cm,width=6cm]{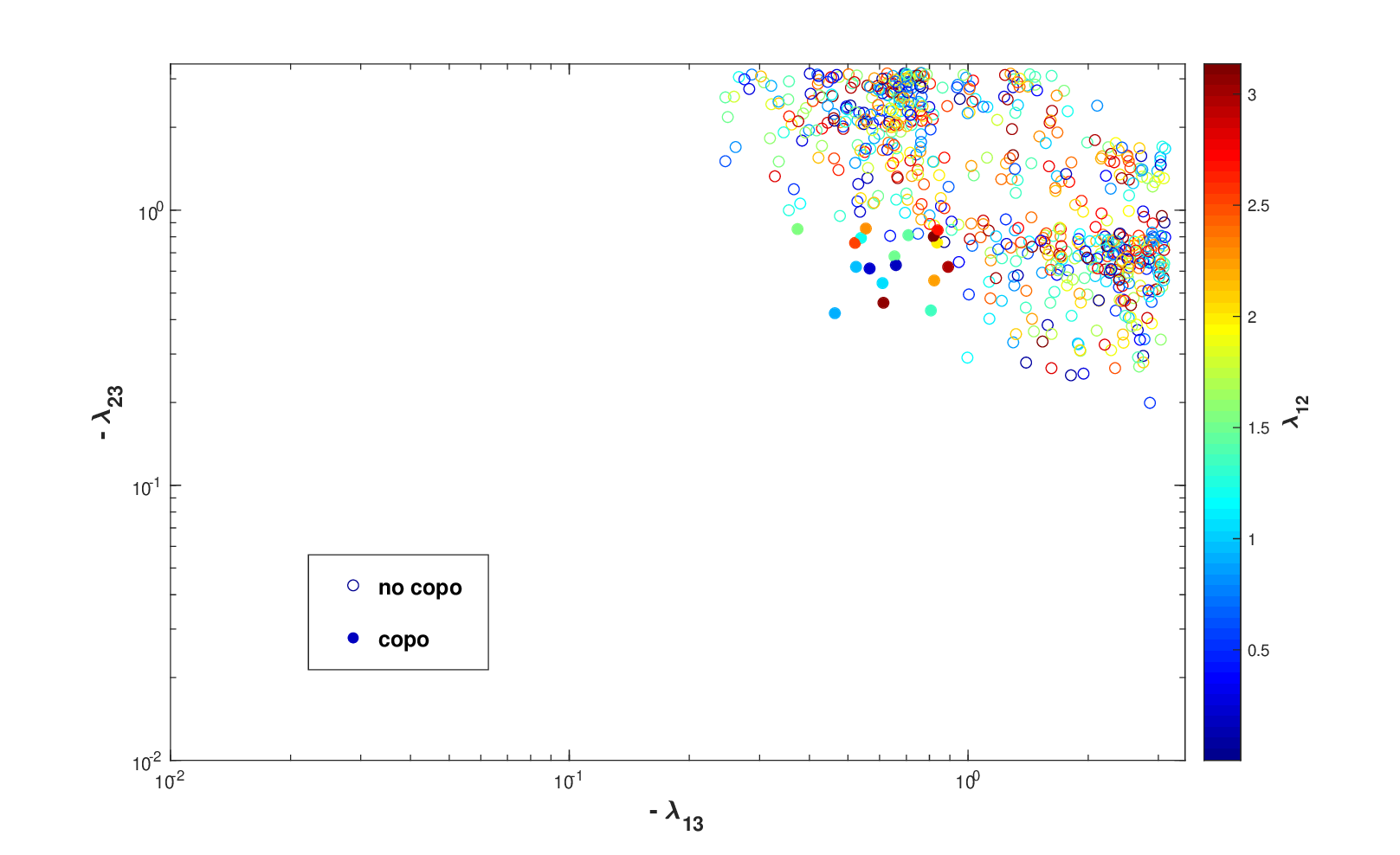}}
\caption{Results of viable parameter space  satisfying the relic density and the direct detection constraint for case (4). In (b),(c),(d), the colored points satisfying copositive criterial while the colored bubbles are not, which are labeled by "copo" and "no copo".}
\label{figs8co}
\end{figure}
\subsection{Random scan results}
 In this part, we perfrom a random scan on the chosen parameter space with:
\begin{eqnarray*}
 m_{1,2} \subset [1\mathrm{TeV}, 4\mathrm{TeV}],|\lambda_{12}| \subset [10^{-4},3.14], |\lambda_{13}|\subset [10^{-4},3.14],
 \end{eqnarray*}
 \begin{eqnarray*}
 |\lambda_{14}| \subset [10^{-4},3.14], |\lambda_{23}|\subset [10^{-4},3.14], |\lambda_{24}| \subset [10^{-4},3.14].
 \end{eqnarray*}
while we fix $\sin\theta=0.01, m_{h_2}=1.5$ TeV and $v_1=2$ TeV for simplicity. We consider the copositive criteria, dark matter relic density  and direct detection constraint on the model for the four different cases, and the reuslts are given in Fig.~\ref{figs1co} to  Fig.~\ref{figs8co}.
 
 For the case (1), the copositive criterial makes little difference in the parameter space, and we show the  viable parameter space  satisfying the relic density and the direct detection constraint in Fig.~\ref{figs1co}. Both $m_1$ and $m_2$ can take values ranging from [1~TeV, 4~TeV] as we can see in Fig.~\ref{figs1co}(a). According to (b) and (c), a larger $\lambda_{14}$ ($\lambda_{24}$) can always contribute to a stronger interaction rate so that the fraction of $S_1$ ($S_2$) is smaller. The direct detection constraint limits a small $\lambda_{14}$  and $\lambda_{24}$,  and a large $\lambda_{13}$ as well as $\lambda_{23}$ are  hence demanded under relic density constraint. As we can see in Fig.~\ref{figs1co}(d), most of the points lie in the upper right region for $(\lambda_{13},\lambda_{23})$. In addition, $\lambda_{12}$ is related to the conversion  processes between $S_1$ and $S_2$, which is less constrained.
 
 For the case (2), case (3) and case (4), we consider  the copositive criterial on the parameter space, and we show the viable parameter space  satisfying the relic density and the direct detection constraint in Fig.~\ref{figs4co}, Fig.~\ref{figs6co} and Fig.~\ref{figs8co}, where
the colored points satisfying copositive criterial while the colored bubbles are not, which are labeled by "copo" and "no copo". We have similar conclusion for the case (1), but when considering copositive criterial, the parameter space is more constrained and only a few points "survived" according to Fig.~\ref{figs4co} to Fig.~\ref{figs8co}. In addition, for the different copositive criterials, the allowed parameter space are different.

\bibliography {v1}
\end{document}